\documentclass[journal]{IEEEtran}

\usepackage{amsthm}
\usepackage{float}
\usepackage{cite}										
\usepackage{amsmath}						
\usepackage{graphicx}
\usepackage{epstopdf}
\usepackage{float}
\usepackage{bm,upgreek}
\usepackage{amsfonts}
\usepackage{enumitem}
\usepackage{mathtools}
\usepackage{multirow}
\usepackage{booktabs}
\usepackage[subnum]{cases}
\usepackage{setspace}
\usepackage{color}
\usepackage{blkarray, bigstrut} 
\usepackage[caption=false,labelformat=simple]{subfig}

\usepackage{xcolor}
\usepackage{pgfplots}
\pgfplotsset{compat=1.5}
\usepackage{pgfplotstable}
\usepgfplotslibrary{statistics}
\pgfplotsset{grid style={dotted,gray}}
\usepackage{tikz}
\usetikzlibrary{math}
\usepackage{hyperref}
\usepackage{easybmat}

\usepackage{mathtools,nccmath}
\usepackage[ruled]{algorithm}
\usepackage{algpseudocode,algorithmicx}

\algnewcommand{\LineComment}[1]{\State \(\#\) #1}

\allowdisplaybreaks

\ifCLASSINFOpdf
\else
\fi
\pgfplotsset{legend image with text/.style={legend image code/.code={%
\node[anchor=west, align=right] at (0.0cm,0cm) {#1};}},}

\makeatletter
\def\myline{\pgfutil@ifnextchar[{\my@line}{\my@line[]}}%
\def\my@line[#1](#2)(#3){%
\tikz[overlay] \draw[#1]  (#2)--(#3); 
}%
\algrenewcommand\algorithmicindent{1.0em}%

\newtheorem{example}{Example}

\makeatletter
\renewcommand{\ALG@beginalgorithmic}{\small}
\makeatother



\usepackage{etoolbox}
\BeforeBeginEnvironment{algorithmic}{\kern 0.3\baselineskip}
\AfterEndEnvironment{algorithmic}{\kern -0.1\baselineskip}

\usepackage{balance}

\makeatletter
\renewcommand{\ALG@beginalgorithmic}{\footnotesize}
\makeatother

\pgfplotsset{
    box plot/.style={
        /pgfplots/.cd,
        black,
        only marks,
        mark=-,
        mark size=\pgfkeysvalueof{/pgfplots/box plot width},
        /pgfplots/error bars/y dir=plus,
        /pgfplots/error bars/y explicit,
        /pgfplots/table/x index=\pgfkeysvalueof{/pgfplots/box plot x index},
    },
    box plot box/.style={
        /pgfplots/error bars/draw error bar/.code 2 args={%
            \draw  ##1 -- ++(\pgfkeysvalueof{/pgfplots/box plot width},0pt) |- ##2 -- ++(-\pgfkeysvalueof{/pgfplots/box plot width},0pt) |- ##1 -- cycle;
        },
        /pgfplots/table/.cd,
        y index=\pgfkeysvalueof{/pgfplots/box plot box top index},
        y error expr={
            \thisrowno{\pgfkeysvalueof{/pgfplots/box plot box bottom index}}
            - \thisrowno{\pgfkeysvalueof{/pgfplots/box plot box top index}}
        },
        /pgfplots/box plot
    },
    box plot top whisker/.style={
        /pgfplots/error bars/draw error bar/.code 2 args={%
            \pgfkeysgetvalue{/pgfplots/error bars/error mark}%
            {\pgfplotserrorbarsmark}%
            \pgfkeysgetvalue{/pgfplots/error bars/error mark options}%
            {\pgfplotserrorbarsmarkopts}%
            \path ##1 -- ##2;
        },
        /pgfplots/table/.cd,
        y index=\pgfkeysvalueof{/pgfplots/box plot whisker top index},
        y error expr={
            \thisrowno{\pgfkeysvalueof{/pgfplots/box plot box top index}}
            - \thisrowno{\pgfkeysvalueof{/pgfplots/box plot whisker top index}}
        },
        /pgfplots/box plot
    },
    box plot bottom whisker/.style={
        /pgfplots/error bars/draw error bar/.code 2 args={%
            \pgfkeysgetvalue{/pgfplots/error bars/error mark}%
            {\pgfplotserrorbarsmark}%
            \pgfkeysgetvalue{/pgfplots/error bars/error mark options}%
            {\pgfplotserrorbarsmarkopts}%
            \path ##1 -- ##2;
        },
        /pgfplots/table/.cd,
        y index=\pgfkeysvalueof{/pgfplots/box plot whisker bottom index},
        y error expr={
            \thisrowno{\pgfkeysvalueof{/pgfplots/box plot box bottom index}}
            - \thisrowno{\pgfkeysvalueof{/pgfplots/box plot whisker bottom index}}
        },
        /pgfplots/box plot
    },
    box plot median/.style={
        /pgfplots/box plot,
        /pgfplots/table/y index=\pgfkeysvalueof{/pgfplots/box plot median index},
        thin,black
    },
    box plot width/.initial=1em,
    box plot x index/.initial=0,
    box plot median index/.initial=1,
    box plot box top index/.initial=2,
    box plot box bottom index/.initial=3,
    box plot whisker top index/.initial=4,
    box plot whisker bottom index/.initial=5,
}

\newcommand{\boxplot}[2][]{
    \addplot [box plot median,#1] table {#2};
    \addplot [forget plot, box plot box,#1] table {#2};
    \addplot [forget plot, box plot top whisker,#1] table {#2};
    \addplot [forget plot, box plot bottom whisker,#1] table {#2};
}

\pgfplotsset{my legend/.style={
    legend image code/.code={
        \fill [#1] (0cm,-0.1cm) rectangle (0.3cm,0.1cm);
    },
}}

\algnewcommand\algorithmicforeach{\textbf{for each}}
\algdef{S}[FOR]{ForEach}[1]{\algorithmicforeach\ #1\ \algorithmicdo}

\definecolor{coal}{HTML}{56CBF4}
\definecolor{oil}{HTML}{4876BB}
\definecolor{gas}{HTML}{86CA88}
\definecolor{nuclear}{HTML}{15AEA1}
\definecolor{hydro}{HTML}{F9F162}
\definecolor{geothermal}{HTML}{4F4D2E}
\definecolor{wind}{HTML}{F3744E}
\definecolor{solar}{HTML}{A186BE}
\definecolor{biofuels}{HTML}{AB6AAC}

\begin{document}

\newlist{myitemize}{itemize}{3}
\setlist[myitemize,1]{label=\textbullet,leftmargin=6.5mm}

\title{Vectorized Gaussian Belief Propagation \\ for Near Real-Time Fully-Distributed \\ PMU-Based State Estimation}

\author{Mirsad Cosovic,
        Armin Teskeredzic,
        Antonello Monti,
        Dejan Vukobratovic 

\thanks{M. Cosovic is with Faculty of Electrical Engineering, University of Sarajevo, Bosnia and Herzegovina, and the Institute for Artificial Intelligence Research and Development of Serbia (e-mail: mcosovic@etf.unsa.ba); A. Teskeredzic is with Institute for Automation of Complex Power Systems, E.ON Energy Research Center, RWTH Aachen University, Germany (e-mail: armin.teskeredzic@eonerc.rwth-aachen.de); A. Monti is with the Institute for Automation of Complex Power Systems, E.ON Energy Research Center, RWTH Aachen University, Germany, and with the Fraunhofer FIT, Aachen, Germany (e-mail: amonti@eonerc.rwth-aachen.de) D. Vukobratovic is with Faculty of Technical Sciences, University of Novi Sad, Serbia (email: dejanv@uns.ac.rs).}}

\markboth{}%
{Shell \MakeLowercase{\textit{et al.}}: Bare Demo of IEEEtran.cls for IEEE Journals}

\maketitle

\begin{abstract}
Electric power systems require accurate, scalable, distributed, and near real-time state estimation (SE) to support reliable monitoring and control under increasingly complex operating conditions. Limited monitoring capabilities can lead to inefficient operation and, in extreme cases, large-scale disturbances such as blackouts. To address these challenges, this paper proposes a vectorized Gaussian belief propagation (GBP) framework for phasor measurement unit-based SE, formulated over factor graphs and specifically designed to support distributed and near real-time monitoring. The proposed framework includes multivariate and fusion-based GBP formulations. The multivariate formulation jointly models related state variables and their measurement relationships, while the fusion-based formulation reduces factor graph complexity by combining multiple measurements associated with the same set of variables, resulting in a structure that more closely reflects the underlying electrical coupling of the power system. The resulting algorithms operate in a fully distributed manner at the bus level and achieve fast convergence and high estimation accuracy, often within a few iterations, as demonstrated by numerical results on systems ranging from 60 to 13659 buses, where the fusion-based formulation achieves single-digit millisecond iteration times on the largest test case. 
\end{abstract}


\IEEEpeerreviewmaketitle

\section{Introduction}
Electric power systems constitute critical infrastructure that supports modern society and enables technological and industrial development. The deregulation of energy markets and the integration of renewable energy sources have significantly increased system complexity, exposing the limitations of traditional operational practices. As a result, power systems have become more susceptible to inefficient operation and large-scale disturbances, including blackouts \cite{gonzalez2025challenges}. Efficient monitoring has therefore become essential for reliable system control and must be implemented in a distributed manner while supporting near real-time operation \cite{korres, cosovic20175g}.

The transition toward near real-time monitoring has shifted the focus from legacy measurements with low sampling rates provided by supervisory control and data acquisition systems to phasor measurement units (PMUs), which generate high-rate synchronized measurements. Within wide-area measurement systems, PMU data enable more accurate and timely observation of system states \cite{realtimemonitoring}. In the future, PMU-based monitoring is expected to serve as the foundation for advanced control and protection strategies in modern power systems \cite{pmufuture}. However, the practical value of high-rate PMU data for near real-time monitoring depends on how efficiently these measurements can be processed and translated into reliable estimates of system states.

In this context, the state estimation (SE) algorithm forms the core of modern monitoring systems by estimating the complex bus voltages, which constitute the state variables, from synchronized phasor measurements of voltage and current magnitudes and angles. The SE algorithm must provide accurate and timely state estimates while supporting efficient, robust, and scalable computation. To meet the demands of contemporary power systems, SE must also preserve data privacy among participating entities. Although many studies address either distributed SE \cite{korres, nasiri2023secure, aburpmumulti, marelliwlsdistributed, zhou2020gradient} or near real-time operation \cite{realtimemonitoring, cavraro2022real, chen2016synchrophasor, liu2021dynamic}, only a limited number of works explicitly combine both aspects to achieve simultaneous distributed and near real-time performance \cite{kundacina2022near, wang2020distributed}. In \cite{kundacina2022near}, the approach relies on AI/ML models that require prior training and do not inherently guarantee convergence to the exact solution of the SE problem, which increases model complexity and may limit scalability across varying system conditions. In \cite{wang2020distributed}, the method relies on a consensus+innovations-based distributed estimation process over switching inter-area communication graphs, which requires iterative inter-area information exchange and communication coordination. This may increase communication overhead, make real-time responsiveness dependent on network conditions, and raise data privacy concerns due to the exchange of intermediate estimation information.

To address these challenges, Gaussian belief propagation (GBP) has emerged as a promising approach for SE in power systems. In recent years, GBP has attracted significant attention in various SE-related tasks, including SE algorithms \cite{cosovictra, sun2024complex, yang2020109091}, observability analysis \cite{cosovicobs, ren2022power}, bad data detection \cite{cosovictra}, and cyber threat analysis \cite{wei2026gaussian}. The majority of these approaches rely on scalar or component-wise formulations of GBP defined over factor graphs. In such formulations, scalar decomposition of state variables and phasor measurements introduces a large number of short loops in the factor graph, which can slow convergence and degrade the stability of GBP. Although \cite{sun2024complex} adopts a non-scalar formulation in the complex domain, it relies on variance approximations, which may degrade the accuracy and convergence properties of the SE solution. Moreover, such formulations fail to explicitly model the correlation between the real and imaginary components arising from the transformation of phasor measurements from polar to rectangular coordinates, as required for the linear SE model. In contrast, \cite{yang2020109091} proposes a vectorized GBP approach defined over a variable graph, where measurement relations are implicitly embedded within the edges. Compared to a factor-graph representation, this abstraction reduces structural granularity and makes it more difficult to explicitly track the contribution of individual measurements and to update them efficiently, which complicates the efficient incorporation and updating of measurements in near real-time settings.

Motivated by these limitations, we propose a vectorized GBP framework over factor graphs for PMU-based SE, comprising multivariate and fusion-based formulations. The multivariate GBP jointly models related state variables and measurements, reducing the number of loops in the factor graph compared to the scalar formulation. This representation also preserves the explicit measurement structure inherent in factor-graph models, unlike variable-graph-based approaches. The fusion-based GBP further reduces graph redundancy by combining multiple measurements associated with the same set of state variables, effectively eliminating local loops between pairs of variables. This progressive simplification of the graph structure improves convergence and accelerates the estimation process, enabling faster and more accurate SE, which makes the approach suitable for near real-time monitoring. More precisely, the proposed framework results in a fully distributed algorithm operating at the bus level and achieving high estimation accuracy within a few iterations, as demonstrated in the numerical results. In addition, the message-passing structure of GBP naturally enables the exchange of information in the form of beliefs rather than raw state variables and measurement data, which is beneficial from a data privacy perspective. The proposed framework also does not require centralized coordination or strict synchronization and naturally supports asynchronous operation.

\section{Linear SE in PMU-Observable Power Systems}
In PMU-observable power systems, the SE problem is formulated as a linear model that relates the state variables to phasor measurements through the network topology and electrical parameters of buses and branches. A representation of the network topology is the bus/branch model, described by a graph $\mathcal{P} = (\mathcal{N}, \mathcal{L})$, where nodes $\mathcal{N}$ correspond to buses and edges $\mathcal{L} \subseteq \mathcal{N} \times \mathcal{N}$ represent branches. Each branch is modeled as a unified two-port $\pi$ model with its electrical parameters \cite{andersson2008modelling}. The state variables correspond to complex bus voltages and are represented in rectangular coordinates to obtain a linear measurement model. For each bus $i \in \mathcal{N}$, the local state vector is defined as $\mathbf{x}_i = [V_{\mathrm{re},i}, V_{\mathrm{im},i}]^T$. Collecting these vectors over all $n$ buses yields the set of local state vectors $\mathcal{X} = \{\mathbf{x}_1, \mathbf{x}_2, \dots, \mathbf{x}_n\}$ to be estimated.

The local state vectors in $\mathcal{X}$ are estimated from bus voltage and branch current phasor measurements collected in the set $\mathcal{F} = \{f_1, \dots, f_m\}$. These measurements are provided by PMUs in polar coordinates and transformed into rectangular coordinates in order to obtain a linear measurement model with respect to the state variables~\cite{gomez2011use}:
\begin{equation}
    \mathbf{z}_{f_k} = \begin{bmatrix}
        z_\mathrm{re} \\ z_\mathrm{im}
    \end{bmatrix}
    =
    \begin{bmatrix}
        z_\mathrm{m} \cos z_\uptheta \\ z_\mathrm{m} \sin z_\uptheta
    \end{bmatrix},
	\label{to_rectangular_mean}
\end{equation}
where $z_\mathrm{m}$ and $z_\uptheta$ denote the measured magnitude and angle of phasor measurement $f_k \in \mathcal{F}$, respectively. This transformation also maps the measurement uncertainties into rectangular coordinates, inherently introducing statistical correlation between the real and imaginary components. Following the classical theory of uncertainty propagation \cite{iso1993guide}, the resulting variances and covariance are given by:
\begin{equation}
    \begin{aligned}
        \sigma_\mathrm{re}^2  &= \sigma_\mathrm{m}^2 \cos^2 z_\uptheta + \sigma_\uptheta^2 z_\mathrm{m}^2 \sin^2 z_\uptheta \\
        \sigma_\mathrm{im}^2  &= \sigma_\mathrm{m}^2 \sin^2 z_\uptheta + \sigma_\uptheta^2 z_\mathrm{m}^2 \cos^2 z_\uptheta \\
        \sigma_\mathrm{re,im} &= (\sigma_\mathrm{m}^2 - z_\mathrm{m}^2 \sigma_\uptheta^2)\sin z_\uptheta \cos z_\uptheta,
    \end{aligned}
	\label{to_rectangular_variance}
\end{equation}
where $\sigma_\mathrm{m}^2$ and $\sigma_\uptheta^2$ denote the magnitude and angle variances of phasor measurement $f_k \in \mathcal{F}$. Each phasor measurement is therefore characterized by the non-diagonal covariance matrix:
\begin{equation}
    \mathbf{\Sigma}_{f_k} =
    \begin{bmatrix}
        \sigma_\mathrm{re}^2  & \sigma_\mathrm{re,im} \\
        \sigma_\mathrm{re,im} & \sigma_\mathrm{im}^2
    \end{bmatrix}.
	\label{covariance_matrix}
\end{equation}
The corresponding precision matrix is given by $\mathbf{\Lambda}_{f_k} = \mathbf{\Sigma}_{f_k}^{-1}$, and the covariance and precision representations are used interchangeably throughout the paper.

For a voltage phasor measurement at bus $i \in \mathcal{N}$, the corresponding measurement model is expressed as a local multivariate Gaussian likelihood:
\begin{equation}
    p(\mathbf{z}_{f_k}|\mathbf{x}_i) \propto
    \exp \left( -\frac{1}{2} \| \mathbf{z}_{f_k} - \mathbf{H}_{f_k} \mathbf{x}_i \|^2_{\mathbf{\Lambda}_{f_k}} \right),
	\label{voltage_likelihood}
\end{equation}
where $\mathbf{H}_{f_k} \in \mathbb{R}^{2 \times 2}$ is the identity matrix.

In the case of current phasor measurements on branch $(i,j) \in \mathcal{L}$, the multivariate Gaussian likelihood captures the coupling between buses $i$ and $j$:
\begin{equation}
    \resizebox{0.91\columnwidth}{!}{%
    $p(\mathbf{z}_{f_k}|\mathbf{x}_i, \mathbf{x}_j) \propto
    \exp \left( -\frac{1}{2} \left\| \mathbf{z}_{f_k} -
    \begin{bmatrix} \mathbf{H}_{f_k,\mathbf{x}_i} \; \mathbf{H}_{f_k,\mathbf{x}_j} \end{bmatrix}
    \begin{bmatrix} \mathbf{x}_i \\ \mathbf{x}_j \end{bmatrix} \right\|^2_{\mathbf{\Lambda}_{f_k}} \right),$%
    }
	\label{current_likelihood}
\end{equation}
where the submatrices $\mathbf{H}_{f_k,\mathbf{x}_i}, \mathbf{H}_{f_k,\mathbf{x}_j} \in \mathbb{R}^{2 \times 2}$ are determined by the measurement direction and the electrical parameters of the corresponding branch. When the current is measured at bus $i$ toward bus $j$, these matrices are:
\begin{equation}
    \mathbf{H}_{f_k,\mathbf{x}_i} = \frac{1}{\tau_{ij}^2}(\mathbf{Y}_{ij} + \mathbf{Y}_{\mathrm{s}ij}),\;\;\;
    \mathbf{H}_{f_k,\mathbf{x}_j} = - \frac{1}{\tau_{ij}} \mathbf{Y}_{ij} \mathbf{R}(\phi_{ij}),
\end{equation}
while for measurements at bus $j$ toward bus $i$:
\begin{equation}
    \mathbf{H}_{f_k,\mathbf{x}_i} = -\frac{1}{\tau_{ij}} \mathbf{Y}_{ij} \mathbf{R}(-\phi_{ij}),\;\;\;
    \mathbf{H}_{f_k,\mathbf{x}_j} = \mathbf{Y}_{ij} + \mathbf{Y}_{\mathrm{s}ij}.
\end{equation}
The matrices $\mathbf{Y}_{ij}$, $\mathbf{Y}_{\mathrm{s}ij}$, and $\mathbf{R}(\phi_{ij})$ are given by:
\begin{align}
    \mathbf{Y}_{ij} &=
    \begin{bmatrix}
        g_{ij} & - b_{ij} \\ b_{ij} & g_{ij}
    \end{bmatrix}, \;\;\;
    \mathbf{Y}_{\mathrm{s}ij} =
    \begin{bmatrix}
        g_{\mathrm{s}ij} & - b_{\mathrm{s}ij} \\ b_{\mathrm{s}ij} & g_{\mathrm{s}ij}
    \end{bmatrix}
    \label{Ymatrices} \\
    &\mathbf{R}(\phi_{ij}) =
    \begin{bmatrix}
        \cos \phi_{ij} & - \sin \phi_{ij} \\ \sin \phi_{ij} & \cos \phi_{ij}
    \end{bmatrix}, \nonumber
\end{align}
where $g_{ij}$ and $b_{ij}$ are the series conductance and susceptance, $g_{\mathrm{s}ij}$ and $b_{\mathrm{s}ij}$ are the shunt conductance and susceptance, and $\tau_{ij}$ and $\phi_{ij}$ are the transformer tap ratio and phase shift angle of branch $(i,j) \in \mathcal{L}$.

The solution to the SE problem is obtained by maximizing the joint likelihood function constructed from $m$ independent phasor measurements:
\begin{equation}
    \hat{\mathbf{x}} = \arg \max_\mathbf{x} \prod_{f_k \in \mathcal{V}}
    p(\mathbf{z}_{f_k}|\mathbf{x}_i) \prod_{f_k \in \mathcal{I}} p(\mathbf{z}_{f_k}|\mathbf{x}_i, \mathbf{x}_j),
	\label{likelihood_solution}
\end{equation}
where $\mathcal{V}$ and $\mathcal{I}$ denote the sets of voltage and current phasor measurements, respectively, such that $\mathcal{F} = \mathcal{V} \cup \mathcal{I}$. The global state vector $\mathbf{x} \in \mathbb{R}^{2n}$ is formed by stacking all local state vectors from the set $\mathcal{X}$.

The solution in \eqref{likelihood_solution} is equivalent to the weighted least-squares (WLS) estimator~\cite[Ch.~2]{abur}:
\begin{equation}
    \hat{\mathbf{x}} = (\mathbf{H}^T \mathbf{\Sigma}^{-1} \mathbf{H})^{-1} \mathbf{H}^T \mathbf{\Sigma}^{-1} \mathbf{z},
	\label{wls_solution}
\end{equation}
where the individual measurements $\mathbf{z}_{f_k}$ are concatenated into the global measurement vector $\mathbf{z} \in \mathbb{R}^{2m}$. The global measurement matrix $\mathbf{H} \in \mathbb{R}^{2m \times 2n}$ and the block-diagonal covariance matrix $\mathbf{\Sigma} \in \mathbb{R}^{2m \times 2m}$ are assembled from the local submatrices $\mathbf{H}_{f_k}$, $\mathbf{H}_{f_k,\mathbf{x}_i}$, $\mathbf{H}_{f_k,\mathbf{x}_j}$, and $\mathbf{\Sigma}_{f_k}$. The matrix $\mathbf{H}$ is assumed to have full column rank, which ensures full observability under the deployed PMU configuration.

The estimator in \eqref{wls_solution} is typically implemented in centralized SE approaches, whereas distributed approaches rely on suitable decomposition or reformulation of the estimation problem. The multivariate GBP formulation, introduced in the next section, operates directly on the vector measurement models and solves \eqref{likelihood_solution}, providing a computationally efficient estimator suitable for near real-time applications. By contrast, the scalar formulation decomposes each multivariate Gaussian likelihood into individual univariate likelihoods. This decomposition increases the number of loops in the graph and may degrade convergence behavior, resulting in slower convergence or even divergence compared to the multivariate formulation.

\section{Multivariate GBP Algorithm}
The GBP algorithm operates by iteratively passing messages over a factor graph composed of variable and factor nodes. According to \eqref{likelihood_solution}, the variable nodes are given by the set $\mathcal{X}$, where each node corresponds to a local state vector $\mathbf{x}_i \in \mathcal{X}$ and can be interpreted as representing a bus. The factor nodes are induced by the likelihood functions in \eqref{likelihood_solution} and associated with the set $\mathcal{F}$, where each factor node corresponds to a phasor measurement $f_k \in \mathcal{F}$. Together, these nodes define the factor graph $\mathcal{G} = (\mathcal{X} \cup \mathcal{F}, \mathcal{E})$, where edges in $\mathcal{E}$ connect variable and factor nodes. In this graph, factor nodes associated with voltage measurements, $p(\mathbf{z}_{f_k}|\mathbf{x}_i)$, connect to variable node $\mathbf{x}_i$, while factor nodes associated with current measurements, $p(\mathbf{z}_{f_k}|\mathbf{x}_i, \mathbf{x}_j)$, connect to both $\mathbf{x}_i$ and $\mathbf{x}_j$. The set of factor nodes $\mathcal{F}$ is therefore partitioned into unary and pairwise factors, corresponding to the sets $\mathcal{V}$ and $\mathcal{I}$, respectively.

The factor graph corresponding to the proposed multivariate formulation exhibits significantly lower complexity than its scalar counterpart. In the scalar formulation, each vector variable node $\mathbf{x}_i \in \mathcal{X}$ is decomposed into two scalar variable nodes. Similarly, each pairwise factor node $f_k \in \mathcal{I}$ is decomposed into two scalar factor nodes, each connected to four scalar variables. As a result, a single branch current phasor measurement requires a total of eight edges in the scalar graph. For voltage phasor measurements, neglecting covariance results in two unary factor nodes connected by two edges, whereas preserving covariance requires four edges and introduces additional loops in the graph. Consequently, the multivariate formulation yields a more compact graph representation, reduces the computational overhead of message passing, and preserves the complete measurement model.

\begin{example}[Constructing a factor graph] 
In this toy example, we consider a 3-bus system in which each branch $(i,j) \in \mathcal{L}$ is characterized by a resistance $r_{ij} = 0.1\,\text{p.u.}$ and a reactance $x_{ij} = 0.2\,\text{p.u.}$. The corresponding bus/branch model, with PMUs installed at buses 1 and 2, is shown in \figurename~\ref{example1_bus_branch}. The corresponding measurement values in polar coordinates are summarized in Table~\ref{pmu_measurements}. For simplicity, all magnitude measurements are assumed to have variance $\sigma_\mathrm{m}^2 = 10^{-6}$, and all angle measurements are assumed to have variance $\sigma_\uptheta^2 = 10^{-4}$. These measurements are used in the following examples to illustrate the GBP message-passing procedure.
\begin{table}[ht]
	\centering
	\vspace{2pt} 
	\caption{Phasor measurements in polar coordinates}
	\label{pmu_measurements}
	\footnotesize
	\setlength{\tabcolsep}{4pt}
	\begin{tabular}{l@{\hspace{15pt}}l r r}
		\midrule
		Phasor measurement & $f_k$ & $z_\mathrm{m}$ (p.u.) & $z_\uptheta$ (rad) \\
		\midrule
		Voltage at bus 1 & $f_1$ & 1.11 & -0.10 \\
		Voltage at bus 2 & $f_2$ & 0.94 & -0.11 \\
		Current from bus 1 to bus 2 & $f_3$ & 0.71 & -1.22 \\
		Current from bus 1 to bus 3 & $f_4$ & 0.67 & -2.01 \\
		Current from bus 2 to bus 1 & $f_5$ & 0.71 & 1.92 \\
		Current from bus 2 to bus 3 & $f_6$ & 0.53 & 3.03 \\
		\midrule
	\end{tabular}
	\vspace{2pt} 
\end{table}

Based on the bus/branch model and the corresponding measurement configuration, the factor graph is constructed as shown in \figurename~\ref{example1_factor_graph}.

\begin{figure}[ht]
	\centering \captionsetup[subfigure]{oneside,margin={0.1cm,0cm}}
	\begin{tabular}{@{\hspace{0cm}}c@{\hspace{0cm}}} \subfloat[]{\label{example1_bus_branch}
	\includegraphics[width=3.2cm]{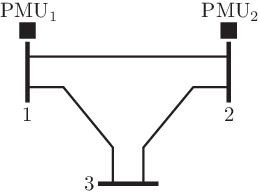}}
	\end{tabular}\quad
	\begin{tabular}{@{\hspace{0cm}}c@{\hspace{0cm}}} \subfloat[]{\label{example1_factor_graph}
	\includegraphics[width=4cm]{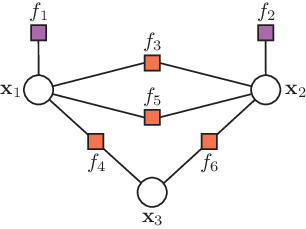}} 
	\end{tabular}
	\caption{Transformation of the bus/branch model and measurement configuration (subfigure a) into the corresponding factor graph with different types of factor nodes (subfigure b).}
\label{example1}
\end{figure} 

The set of variable nodes is given by $\mathcal{X} = \{\mathbf{x}_1, \mathbf{x}_2, \mathbf{x}_3\}$. The PMUs installed at buses 1 and 2 measure the voltage phasors at their respective buses, giving rise to the unary factor nodes $\mathcal{V} = \{f_1, f_2\}$ (purple boxes). In addition, these PMUs measure branch currents at their bus locations. The PMU at bus 1 measures currents flowing toward buses 2 and 3, while the PMU at bus 2 measures currents flowing toward buses 1 and 3. These measurements give rise to the pairwise factor nodes $\mathcal{I} = \{f_3, f_4, f_5, f_6\}$ (orange boxes). Together, they form the complete set of factor nodes, $\mathcal{F} = \mathcal{V} \cup \mathcal{I}$.
\end{example}

The GBP algorithm iteratively exchanges messages between variable nodes $\mathbf{x}_i \in \mathcal{X}$ and pairwise factor nodes $f_k \in \mathcal{I}$. Due to linearity, each message follows a multivariate Gaussian distribution and is therefore represented by a mean vector and a precision matrix. Unary factors $f_k \in \mathcal{V}$ contribute constant messages determined by the measurement mean $\mathbf{z}_{f_k}$ and covariance matrix $\mathbf{\Sigma}_{f_k}$, which remain fixed throughout the iterations. At each iteration, a variable node aggregates incoming messages from all neighboring factors to form its belief, thereby combining information from both unary and pairwise factors. This belief is then used to generate outgoing messages to neighboring pairwise factors, enabling iterative refinement of the state estimates through repeated message updates until convergence.

Prior to the iterative message-passing procedure, the messages from variable nodes $\mathbf{x}_i \in \mathcal{X}$ to pairwise factor nodes $f_k \in \mathcal{I}$ must be initialized. This is typically achieved by propagating the measurement mean and covariance from unary factors $f_k \in \mathcal{V}$ through the corresponding variable nodes toward all connected pairwise factors. For variable nodes without connected unary factors, the messages are initialized using suitable mean values and large variances to reflect the initial uncertainty. These initial values may also incorporate prior knowledge about unmeasured variables, thereby providing a warm start for the algorithm.

\begin{example}[Initialization]
To illustrate the initialization of GBP, we consider the 3-bus system and the corresponding factor graph shown in \figurename~\ref{example1}. The initialization requires defining all messages from variable nodes $\mathbf{x}_i \in \mathcal{X}$ to pairwise factor nodes $f_k \in \mathcal{I}$, as depicted in \figurename~\ref{example2}.

\begin{figure}[ht]
	\centering
    \includegraphics[width=4.1cm]{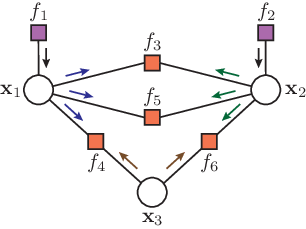}
    \caption{Illustration of the initialization procedure for the GBP algorithm, showing how measurement information and prior estimates are propagated from variable nodes to factor nodes.}
\label{example2}
\end{figure}

Consider the voltage phasor measurement corresponding to factor node $f_1$. After transforming the measurement from polar to rectangular coordinates, the mean is $\mathbf{z}_{f_1} = [1.10,\, -0.11]^T$, with precision matrix $\mathbf{\Lambda}_{f_1} = 10^5 \cdot \mathrm{diag}(4.51,\, 0.08)$, where the covariance is neglected for simplicity in this illustrative example. To initialize the GBP algorithm, these parameters are propagated from the variable node $\mathbf{x}_1$ along all incident edges toward the pairwise factor nodes $f_3$, $f_4$, and $f_5$, as indicated by the blue arrows in the figure. For example, the outgoing message from $\mathbf{x}_1$ to factor node $f_4$ has mean $\mathbf{z}_{\mathbf{x}_1 \to f_4} = \mathbf{z}_{f_1}$ and precision $\mathbf{\Lambda}_{\mathbf{x}_1 \to f_4} = \mathbf{\Lambda}_{f_1}$. Similarly, variable node $\mathbf{x}_2$ receives information from its connected unary factor $f_2$ and propagates the corresponding mean and precision along its incident edges toward factor nodes $f_3$, $f_5$, and $f_6$, as indicated by the green arrows.

For variable node $\mathbf{x}_3$, which has no connected unary factor, the initial mean and covariance are selected to reflect the absence of a voltage measurement. For example, a mean $\mathbf{z}_{\mathbf{x}_3} = [1.0,\, 0.0]^T$ can be used together with a large covariance, corresponding to a precision matrix $\mathbf{\Lambda}_{\mathbf{x}_3} = \mathrm{diag}(10^{-8},\, 10^{-8})$. These parameters are then propagated along the incident edges toward factor nodes $f_4$ and $f_6$, as indicated by the brown arrows in the figure.
\end{example}

\subsection{Message from Factor Node to Variable Node}
\begin{figure}[ht]
	\centering
	\includegraphics[width=5.5cm]{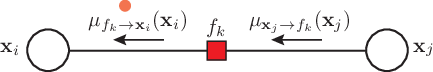}
	\caption{Message $\mu_{f_k \to \mathbf{x}_i}(\mathbf{x}_i)$ from factor node $f_k$ to variable node $\mathbf{x}_i$.}
\label{fg_fac_to_var}
\end{figure}

For a message from a factor node to a variable node, the pairwise structure implies that only one incoming message is available, which simplifies the computation. Without loss of generality, we consider the derivation of the message $\mu_{f_k \to \mathbf{x}_i}(\mathbf{x}_i)$ from the factor node $f_k \in \mathcal{I}$ to the variable node $\mathbf{x}_i \in \mathcal{X}$. The corresponding factor graph is shown in \figurename~\ref{fg_fac_to_var}.

The message $\mu_{f_k \to \mathbf{x}_i}(\mathbf{x}_i)$ is obtained by multiplying the incoming message $\mu_{\mathbf{x}_j \to f_k}(\mathbf{x}_j)$ with the local likelihood function $p(\mathbf{z}_{f_k}|\mathbf{x}_i, \mathbf{x}_j)$ defined in \eqref{current_likelihood}, and marginalising over the variable $\mathbf{x}_j$:
\begin{equation}
    \mu_{f_k \to \mathbf{x}_i}(\mathbf{x}_i) = 
    \int p(\mathbf{z}_{f_k}|\mathbf{x}_i, \mathbf{x}_j)\,
    \mu_{\mathbf{x}_j \to f_k}(\mathbf{x}_j)\,\mathrm{d}\mathbf{x}_j.
\label{FG_f_v}
\end{equation}

Once the incoming message $\mu_{\mathbf{x}_j \to f_k}(\mathbf{x}_j)$ is available, characterised by the mean $\mathbf{z}_{\mathbf{x}_j \to f_k}$ and precision matrix $\mathbf{\Lambda}_{\mathbf{x}_j \to f_k}$, the resulting message $\mu_{f_k \to \mathbf{x}_i}(\mathbf{x}_i)$ is obtained from \eqref{FG_f_v} as a multivariate Gaussian distribution:
\begin{equation}
    \mu_{f_k \to \mathbf{x}_i}(\mathbf{x}_i) \propto 
    \exp \Big( -\frac{1}{2} \| \mathbf{x}_i - 
    \mathbf{z}_{f_k \to \mathbf{x}_i} \|^2_{\mathbf{\Lambda}_{f_k \to \mathbf{x}_i}} \Big),
\label{GBP_incoming_v_to_f}
\end{equation}
with precision matrix $\mathbf{\Lambda}_{f_k \to \mathbf{x}_i}$ and mean vector $\mathbf{z}_{f_k \to \mathbf{x}_i}$:
\begin{subequations}
    \begin{align}
        \mathbf{\Lambda}_{f_k \to \mathbf{x}_i} &= 
        \mathbf{H}_{f_k,\mathbf{x}_i}^T 
        \mathbf{V}_{f_k}^{-1}
        \mathbf{H}_{f_k,\mathbf{x}_i}
		\label{BP_fv_var} \\
		\mathbf{z}_{f_k \to \mathbf{x}_i} &= 
        \mathbf{H}_{f_k,\mathbf{x}_i}^{-1}
        \big(\mathbf{z}_{f_k} - \mathbf{H}_{f_k,\mathbf{x}_j}\mathbf{z}_{\mathbf{x}_j \to f_k}\big),
        \label{BP_fv_mean}
    \end{align}
\label{BP_fv_mean_var}%
\end{subequations}
where the innovation covariance matrix $\mathbf{V}_{f_k}$ is given by:
\begin{equation}
    \mathbf{V}_{f_k} = \mathbf{\Sigma}_{f_k} + \mathbf{H}_{f_k,\mathbf{x}_j}\mathbf{\Lambda}_{\mathbf{x}_j \to f_k}^{-1} 
    \mathbf{H}_{f_k,\mathbf{x}_j}^T.
    \label{sigma_total_def}
\end{equation}
Once \eqref{BP_fv_mean_var} has been evaluated, the factor node $f_k$ sends the resulting message to the variable node $\mathbf{x}_i$. The message $\mu_{f_k \to \mathbf{x}_j}(\mathbf{x}_j)$ is obtained analogously by exchanging the subscripts $\mathbf{x}_i$ and $\mathbf{x}_j$.

\begin{example}[Messages from Factor Nodes to Variable Nodes] 
The GBP algorithm computes all messages $\mu_{f_k \to \mathbf{x}_i}(\mathbf{x}_i)$ from pairwise factor nodes $f_k \in \mathcal{I}$ to variable nodes $\mathbf{x}_i \in \mathcal{X}$ using \eqref{BP_fv_mean_var}, as illustrated in \figurename~\ref{example3}. This requires the incoming messages $\mu_{\mathbf{x}_i \to f_k}(\mathbf{x}_i)$ from variable nodes to pairwise factor nodes, which are available either from the previous iteration or, in the first iteration, from the initialization.

\begin{figure}[ht]
	\centering
	\includegraphics[width=4.1cm]{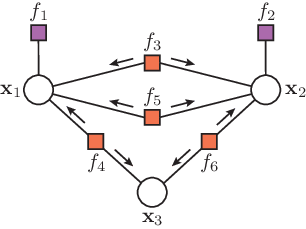}
	\caption{Messages $\mu_{f_k \to \mathbf{x}_i}(\mathbf{x}_i)$ from pairwise factor nodes to variable nodes.}
	\label{example3}
\end{figure}

Consider the current phasor measurement associated with factor node $f_4$, modeled according to \eqref{current_likelihood}. The measurement value is $\mathbf{z}_{f_4} = [-0.28,\, -0.61]^T$, with precision matrix $\mathbf{\Lambda}_{f_4} = 10^5 \cdot \mathrm{diag}(0.27, 1.12)$, and coefficient matrices:
\begin{equation}
    \mathbf{H}_{f_4, \mathbf{x}_1} = 
    \begin{bmatrix*}[r]
          2.0 & 4.0 \\
 		 -4.0 & 2.0
    \end{bmatrix*}, \;\;\;
    \mathbf{H}_{f_4, \mathbf{x}_3} =
    \begin{bmatrix*}[r]
         -2.0 & -4.0 \\
          4.0 & -2.0
    \end{bmatrix*}.
\end{equation}

For illustration, consider the message from factor node $f_4$ to variable node $\mathbf{x}_3$. The incoming message, characterized by the mean $\mathbf{z}_{\mathbf{x}_1 \to f_4}$ and precision matrix $\mathbf{\Lambda}_{\mathbf{x}_1 \to f_4}$, is taken from the initialization. The precision and mean of the outgoing message are then computed using \eqref{BP_fv_mean_var}:
\begin{subequations}
    \begin{align}
     	\mathbf{\Lambda}_{f_4 \to \mathbf{x}_3} &= 
        \mathbf{H}_{f_4,\mathbf{x}_3}^T 
        \mathbf{V}_{f_4}^{-1}
        \mathbf{H}_{f_4,\mathbf{x}_3} \\
		\mathbf{z}_{f_4 \to \mathbf{x}_3} &= 
        \mathbf{H}_{f_4,\mathbf{x}_3}^{-1}
        \big(\mathbf{z}_{f_4} - \mathbf{H}_{f_4,\mathbf{x}_1}\mathbf{z}_{\mathbf{x}_1 \to f_4}\big),
    \end{align}	
\end{subequations}
where the innovation covariance matrix is:
\begin{equation}
    \mathbf{V}_{f_4} = \mathbf{\Sigma}_{f_4} + \mathbf{H}_{f_4,\mathbf{x}_1}\mathbf{\Lambda}_{\mathbf{x}_1 \to f_4}^{-1} \mathbf{H}_{f_4,\mathbf{x}_1}^T.
\end{equation}
The resulting message parameters are:
\begin{equation}
    \mathbf{\Lambda}_{f_4 \to \mathbf{x}_3} = 10^4 \cdot
	\begin{bmatrix*}[r]
		33.98 & -0.15 \\
		-0.15 & 0.81
	\end{bmatrix*}, \;\;
	\mathbf{z}_{f_4 \to \mathbf{x}_3} = 
    \begin{bmatrix*}[r]
    		1.01 \\ 0.01 
    \end{bmatrix*}.
\end{equation}

The remaining messages are computed analogously and used in subsequent iterations to update the messages from variable nodes to pairwise factor nodes.
\end{example}

\subsection{Message from Variable Node to Factor Node}
\begin{figure}[ht]
	\centering
	\includegraphics[width=4.0cm]{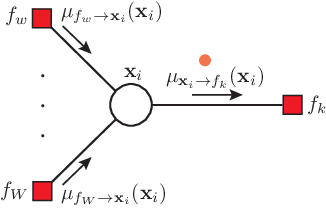}
	\caption{Message $\mu_{\mathbf{x}_i \to f_k}(\mathbf{x}_i)$ from variable node $\mathbf{x}_i$ to factor node $f_k$.}
	\label{fg_var_to_fac}
\end{figure}

Consider the variable node $\mathbf{x}_i \in \mathcal{X}$, which sends the message $\mu_{\mathbf{x}_i \to f_k}(\mathbf{x}_i)$ to a neighboring factor node $f_k \in \mathcal{I}$. The corresponding portion of the factor graph is shown in \figurename~\ref{fg_var_to_fac}, where $\mathcal{F}_i = \{f_k, f_w, \dots, f_W\} \subseteq \mathcal{F}$ denotes the set of factor nodes neighboring $\mathbf{x}_i$.

The message $\mu_{\mathbf{x}_i \to f_k}(\mathbf{x}_i)$ is obtained as the product of all incoming messages from factor nodes in the set $\mathcal{F}_i \setminus \{f_k\}$:
\begin{equation}
    \mu_{\mathbf{x}_i \to f_k}(\mathbf{x}_i) = \prod_{f_a \in \mathcal{F}_i \setminus \{f_k\}} \mu_{f_a \to \mathbf{x}_i}(\mathbf{x}_i),
    \label{FG_v_f}
\end{equation}
where each incoming message $\mu_{f_a \to \mathbf{x}_i}(\mathbf{x}_i)$ is a multivariate Gaussian distribution characterized by the mean vector $\mathbf{z}_{f_a \to \mathbf{x}_i}$ and precision matrix $\mathbf{\Lambda}_{f_a \to \mathbf{x}_i}$. Consequently, the message defined in \eqref{FG_v_f} is also a multivariate Gaussian distribution:
\begin{equation}
    \mu_{\mathbf{x}_i \to f_k}(\mathbf{x}_i) \propto \exp \Big( -\frac{1}{2} \| \mathbf{x}_i - \mathbf{z}_{\mathbf{x}_i \to f_k} \|^2_{\mathbf{\Lambda}_{\mathbf{x}_i \to f_k}} \Big),
	\label{BP_Gauss_vf} 
\end{equation}
with precision matrix $\mathbf{\Lambda}_{\mathbf{x}_i \to f_k}$ and mean vector $\mathbf{z}_{\mathbf{x}_i \to f_k}$:
\begin{subequations}
    \begin{align}
        \mathbf{\Lambda}_{\mathbf{x}_i \to f_k} &= \sum_{f_a \in \mathcal{F}_i \setminus \{f_k\}} \mathbf{\Lambda}_{f_a \to \mathbf{x}_i}
        \label{BP_vf_var} \\
        \mathbf{z}_{\mathbf{x}_i \to f_k} &= \mathbf{\Lambda}_{\mathbf{x}_i \to f_k}^{-1} \sum_{f_a \in \mathcal{F}_i \setminus \{f_k\}} 
        \mathbf{\Lambda}_{f_a \to \mathbf{x}_i} \mathbf{z}_{f_a \to \mathbf{x}_i}.
        \label{BP_vf_mean}
    \end{align}%
	\label{BP_vf_mean_var}%
\end{subequations}
After computing \eqref{BP_vf_mean_var}, the variable node $\mathbf{x}_i$ sends the resulting message to the factor node $f_k$.

\begin{example}[Messages from Variable Nodes to Factor Nodes] 
After computing all messages $\mu_{f_k \to \mathbf{x}_i}(\mathbf{x}_i)$ from pairwise factor nodes to variable nodes, and given the fixed messages from unary factor nodes, the complete set of incoming messages becomes available. These messages are then used to compute all messages $\mu_{\mathbf{x}_i \to f_k}(\mathbf{x}_i)$ from variable nodes $\mathbf{x}_i \in \mathcal{X}$ to pairwise factor nodes $f_k \in \mathcal{I}$ using \eqref{BP_vf_mean_var}, as shown in \figurename~\ref{example4}.

\begin{figure}[ht]
	\centering
	\includegraphics[width=4.1cm]{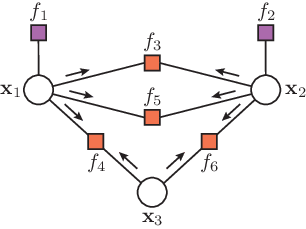}
	\caption{Messages $\mu_{\mathbf{x}_i \to f_k}(\mathbf{x}_i)$ from variable nodes to pairwise factor nodes.}
\label{example4}
\end{figure}

Consider the message from variable node $\mathbf{x}_3$ to factor node $f_6$. The incoming message, characterized by the mean $\mathbf{z}_{f_4 \to \mathbf{x}_3}$ and precision matrix $\mathbf{\Lambda}_{f_4 \to \mathbf{x}_3}$, is obtained from the previous step, in which messages from factor nodes to variable nodes are computed. Since only a single incoming message is present, the message to factor node $f_6$ is equal to that incoming message according to \eqref{BP_vf_mean_var}:
\begin{subequations}
    \begin{align}
    		\mathbf{\Lambda}_{\mathbf{x}_3 \to f_6} &= \mathbf{\Lambda}_{f_4 \to \mathbf{x}_3} \\
        \mathbf{z}_{\mathbf{x}_3 \to f_6} &= 
        \mathbf{\Lambda}_{\mathbf{x}_3 \to f_6}^{-1} 
        \mathbf{\Lambda}_{f_4 \to \mathbf{x}_3} 
        \mathbf{z}_{f_4 \to \mathbf{x}_3} 
        = \mathbf{z}_{f_4 \to \mathbf{x}_3}.
    \end{align}
\end{subequations}

The remaining messages are computed analogously and used in subsequent iterations to update the messages from pairwise factor nodes to variable nodes.
\end{example}

\subsection{Marginal Inference}
\begin{figure}[ht]
	\centering
	\includegraphics[width=3.9cm]{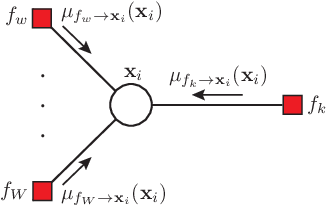}
	\caption{Marginal inference of the variable node $\mathbf{x}_i$.}
	\label{marginal}
\end{figure}

The marginal of the variable node $\mathbf{x}_i \in \mathcal{X}$, illustrated in \figurename~\ref{marginal}, is obtained as the product of all incoming messages to the variable node:
\begin{equation}
    p(\mathbf{x}_i) = \prod_{f_a \in \mathcal{F}_i} \mu_{f_a \to \mathbf{x}_i}(\mathbf{x}_i),
	\label{FG_marginal}
\end{equation}
where $\mathcal{F}_i$ denotes the set of factor nodes incident to the variable node $\mathbf{x}_i$.

From \eqref{FG_marginal}, the marginal of $\mathbf{x}_i$ is also a multivariate Gaussian distribution:
\begin{equation}
    p(\mathbf{x}_i) \propto 
    \exp \left( -\frac{1}{2} \| \mathbf{x}_i - \hat{\mathbf{x}}_i \|^2_{\mathbf{\Lambda}_{\mathbf{x}_i}} \right),
	\label{BP_marginal_gauss_exp}
\end{equation} 
with precision matrix $\mathbf{\Lambda}_{\mathbf{x}_i}$ and mean vector $\hat{\mathbf{x}}_i$ given by:
\begin{subequations}
    \begin{align}
		\mathbf{\Lambda}_{\mathbf{x}_i} &= \sum_{f_a \in \mathcal{F}_i} \mathbf{\Lambda}_{f_a \to \mathbf{x}_i} \\
        \hat{\mathbf{x}}_i &= \mathbf{\Lambda}_{\mathbf{x}_i}^{-1} \sum_{f_a \in \mathcal{F}_i} 
        \mathbf{\Lambda}_{f_a \to \mathbf{x}_i} \mathbf{z}_{f_a \to \mathbf{x}_i}.
    \end{align}
    \label{BP_marginal_mean_var}%
\end{subequations} 
Finally, the mean $\hat{\mathbf{x}}_i$ is taken as the estimate of the local state vector $\mathbf{x}_i$.

\begin{example}[Marginal Inference] 
Marginals can be computed after each iteration, in which case convergence criteria may be applied directly to them, or after convergence when it is monitored through message updates. In either case, once all messages factor nodes to variable nodes are available, the marginals can be computed, as illustrated in \figurename~\ref{example5}.

\begin{figure}[ht]
	\centering
	\includegraphics[width=4.1cm]{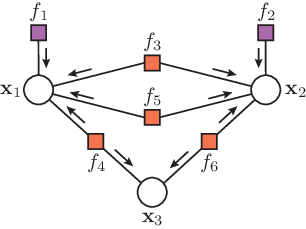}
	\caption{Messages $\mu_{f_k \to \mathbf{x}_i}(\mathbf{x}_i)$ from factor nodes to variable nodes required for marginal computation.}
\label{example5}
\end{figure}

Consider variable node $\mathbf{x}_3$. The incoming messages required to compute its marginal are $\mathbf{z}_{f_4 \to \mathbf{x}_3}$ and $\mathbf{z}_{f_6 \to \mathbf{x}_3}$, with corresponding precision matrices $\mathbf{\Lambda}_{f_4 \to \mathbf{x}_3}$ and $\mathbf{\Lambda}_{f_6 \to \mathbf{x}_3}$. Using \eqref{BP_marginal_mean_var}, the precision matrix and mean vector of the marginal are:
\begin{subequations}
    \begin{align}
    		\mathbf{\Lambda}_{\mathbf{x}_3} &= \mathbf{\Lambda}_{f_4 \to \mathbf{x}_3} + \mathbf{\Lambda}_{f_6 \to \mathbf{x}_3} \\
        \hat{\mathbf{x}}_3 &= \mathbf{\Lambda}_{\mathbf{x}_3}^{-1} \left( \mathbf{\Lambda}_{f_4 \to \mathbf{x}_3} 
        \mathbf{z}_{f_4 \to \mathbf{x}_3} + \mathbf{\Lambda}_{f_6 \to \mathbf{x}_3} \mathbf{z}_{f_6 \to \mathbf{x}_3} \right).
    \end{align}
\end{subequations} 
The resulting marginal parameters are:
\begin{equation}
    \mathbf{\Lambda}_{\mathbf{x}_3} = 10^4 \cdot 
	\begin{bmatrix*}[r] 
    		65.45 & 0.04 \\ 
    		0.04 & 1.95 
	\end{bmatrix*}, \;\;\;
	\hat{\mathbf{x}}_3  = 
    \begin{bmatrix*}[r]
    		1.01 \\ 0.00 
    \end{bmatrix*}.
\end{equation}

The remaining marginals are computed analogously, and the resulting mean vectors are taken as the estimates of the corresponding bus voltages.
\end{example}

\section{Fusion-Based GBP Algorithm}
In power system networks, multiple current phasor measurements may be associated with the same branch, such as measurements available at both ends of a branch or on parallel branches. As a result, multiple pairwise factor nodes may connect to the same set of state variables in the factor graph. Instead of representing these measurements by separate factor nodes, which introduces additional loops, all pairwise factor nodes associated with the same pair of variable nodes are fused into a single equivalent pairwise factor node. This fusion transforms the original set $\mathcal{I}$ into a reduced set of pairwise factor nodes $\widetilde{\mathcal{I}}$. Each factor node $f_k \in \widetilde{\mathcal{I}}$ represents one or more current phasor measurements associated with the same pair of variable nodes, such that each pair of state variables physically connected by a branch is represented by exactly one pairwise factor node. This one-to-one correspondence makes the resulting fused factor graph directly aligned with the observable bus/branch topology defined by the PMU placement. As a result, information propagation in GBP follows the actual electrical couplings, reduces redundant messages, eliminates artificial loops, and improves the convergence properties of the vectorized GBP algorithm.

In the fused representation, the matrices $\mathbf{H}_{f_k,\mathbf{x}_i}$ and $\mathbf{H}_{f_k,\mathbf{x}_j}$ in \eqref{current_likelihood} become rectangular matrices of dimension $2d \times 2$, where $d$ is the number of fused measurements associated with the variables $\mathbf{x}_i$ and $\mathbf{x}_j$. As a result, the message mean update from factor nodes to variable nodes given in \eqref{BP_fv_mean} is no longer directly applicable, since the matrix $\mathbf{H}_{f_k,\mathbf{x}_i}$ is not square and therefore not invertible. In this case, the mean from fused pairwise factor node $f_k \in \widetilde{\mathcal{I}}$ to variable node $\mathbf{x}_i \in \mathcal{X}$ is obtained as:
\begin{equation}
    \mathbf{z}_{f_k \to \mathbf{x}_i} = \mathbf{\Lambda}_{f_k \to \mathbf{x}_i}^{-1} \mathbf{H}_{f_k,\mathbf{x}_i}^T
     \mathbf{V}_{f_k}^{-1} \big(\mathbf{z}_{f_k} - \mathbf{H}_{f_k,\mathbf{x}_j}\mathbf{z}_{\mathbf{x}_j \to f_k}\big).
    \label{fused_mean_message}
\end{equation}
Equation \eqref{BP_fv_mean} is recovered as a special case of \eqref{fused_mean_message}. Therefore, \eqref{fused_mean_message} generalizes the message mean update and remains valid whether the coefficient matrix is square or rectangular. The expression for the precision matrix remains unchanged and is still given by \eqref{BP_fv_var}, indicating that the fusion affects only the mean update.

The primary numerical challenge in this formulation arises from the inversion of the matrix $\mathbf{V}_{f_k}$, whose dimension increases with the number of fused measurements. As a result, this matrix may become ill-conditioned, particularly when there is a significant mismatch between the measurement covariance $\mathbf{\Sigma}_{f_k}$ and the incoming message covariance $\mathbf{\Lambda}_{\mathbf{x}_j \to f_k}^{-1}$. This effect is most pronounced during the initial iterations of the GBP algorithm, when the large variances used for state initialization may dominate the measurement noise. As $\mathbf{\Lambda}_{\mathbf{x}_j \to f_k}^{-1}$ becomes several orders of magnitude larger than $\mathbf{\Sigma}_{f_k}$, the matrix $\mathbf{V}_{f_k}$ may become numerically rank-deficient.

This behavior introduces a trade-off between numerical robustness and convergence efficiency. The scalar GBP formulation is the most robust, since it avoids matrix inversions, followed by the multivariate GBP, while the fused formulation is the most sensitive to ill-conditioning due to the increased dimensionality of $\mathbf{V}_{f_k}$. At the same time, by reducing the number of messages and eliminating artificial loops introduced by parallel edges, the fused factor graph achieves faster convergence than both the multivariate and scalar formulations.

Despite these limitations, the fused approach remains practical. The initial variances can be chosen to represent uncertainty while remaining within the dynamic range of floating-point precision, thereby preserving a reasonable condition number. Moreover, the relatively small size of the fused matrices allows the use of numerically robust factorization methods, such as singular value decomposition, which can handle near-singular cases with minimal impact on the overall computational complexity of the GBP algorithm.

Building on these numerical considerations, the presented formulation provides a theoretically consistent description of the GBP algorithm, but does not fully reflect its most efficient implementation. In practice, the computational complexity can be significantly reduced by exploiting canonical representations and broadcast message updates, further supporting the use of the proposed method for near real-time SE, as detailed in Appendix~A and Appendix~B.

\begin{example}[Messages from Factor Nodes to Variable Nodes with Measurement Fusion] 
Since measurement fusion affects only the computation of messages from factor nodes to variable nodes, we focus on this update. In our example, the pairwise factor nodes $f_3$ and $f_5$ can be fused into a single pairwise factor node $f_7$, since they are incident to the same variable nodes $\mathbf{x}_1$ and $\mathbf{x}_2$, as illustrated in \figurename~\ref{example6}. 

\begin{figure}[ht]
	\centering
	\includegraphics[width=4.1cm]{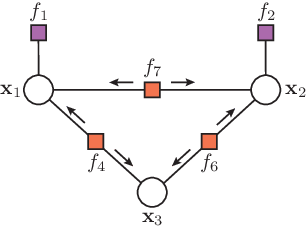}
	\caption{Messages $\mu_{f_k \to \mathbf{x}_i}(\mathbf{x}_i)$ from pairwise factor nodes to variable nodes.}
	\label{example6}
\end{figure}

Consider the fused current phasor measurements associated with factor node $f_7$. The measurement values are stacked into a single vector $\mathbf{z}_{f_7} = [0.24,\, -0.67,\, -0.24,\, 0.67]^T$, with the corresponding precision matrix $\mathbf{\Lambda}_{f_7} = 10^5 \cdot \mathrm{diag}(0.22,\, 1.46,\, 0.22,\, 1.47)$. The associated coefficient matrices are given by:
\begin{equation}
    \scalebox{0.91}{$\mathbf{H}_{f_7, \mathbf{x}_1} = 
    \begin{bmatrix*}[r]
          2.0 & 4.0 \\
 		 -4.0 & 2.0 \\
 		 -2.0 & -4.0  \\
  		  4.0 & -2.0
    \end{bmatrix*}, \;\;\;
    \mathbf{H}_{f_7, \mathbf{x}_2} =
    \begin{bmatrix*}[r]
         -2.0 & -4.0 \\
          4.0 & -2.0 \\
          2.0 &  4.0 \\
         -4.0 &  2.0 \\
    \end{bmatrix*}.$}
\end{equation}

For illustration, consider the message from factor node $f_7$ to variable node $\mathbf{x}_2$. After initialization, the incoming message coincides with the message sent from the unary factor node $f_1$ to $\mathbf{x}_1$, that is, $\mathbf{z}_{\mathbf{x}_1 \to f_7} = \mathbf{z}_{f_1}$ and $\mathbf{\Lambda}_{\mathbf{x}_1 \to f_7} = \mathbf{\Lambda}_{f_1}$. The outgoing message is then computed using \eqref{BP_fv_var} and \eqref{fused_mean_message}:
\begin{subequations}
    \begin{align}
    		\scalebox{0.98}{$\mathbf{\Lambda}_{f_7 \to \mathbf{x}_2}$} &= 
        \scalebox{0.98}{$\mathbf{H}_{f_7,\mathbf{x}_2}^T 
        \mathbf{V}_{f_7}^{-1}
        \mathbf{H}_{f_7,\mathbf{x}_2}$} \\
		\scalebox{0.98}{$\mathbf{z}_{f_7 \to \mathbf{x}_2}$} &= 
        \scalebox{0.98}{$ \mathbf{\Lambda}_{f_7 \to \mathbf{x}_2}^{-1} \mathbf{H}_{f_7,\mathbf{x}_2}^T \mathbf{V}_{f_7}^{-1} 
        \big(\mathbf{z}_{f_7} - \mathbf{H}_{f_7,\mathbf{x}_1}\mathbf{z}_{\mathbf{x}_1 \to f_7}\big),$}
    \end{align}
\end{subequations}
where the innovation covariance matrix is:
\begin{equation}
    \mathbf{V}_{f_7} = \mathbf{\Sigma}_{f_7} + \mathbf{H}_{f_7,\mathbf{x}_1}\mathbf{\Lambda}_{\mathbf{x}_1 \to f_7}^{-1}\mathbf{H}_{f_7,\mathbf{x}_1}^T.
\end{equation}
The resulting message parameters are:
\begin{equation}
    \scalebox{0.91}{$
    \mathbf{\Lambda}_{f_7 \to \mathbf{x}_2} = 10^4 \cdot
	\begin{bmatrix*}[r]
		38.82 & -0.12 \\
		-0.12 & 0.81
	\end{bmatrix*}, \;\;\;
	\mathbf{z}_{f_7 \to \mathbf{x}_2} = 
    \begin{bmatrix*}[r]
    		0.95 \\ -0.09 
    \end{bmatrix*}.$}
\end{equation}

The remaining messages are computed analogously and used in subsequent iterations to update the messages from variable nodes to pairwise factor nodes.
\end{example}

\section{Convergence Analysis of the Fusion-Based GBP Algorithm}
In this section, we analyze the convergence of the fusion-based GBP algorithm. The analysis is focused on the fused formulation, since the multivariate formulation is recovered as a special case for $d = 1$. Although the derivation follows steps similar to those in \cite{du2018convergence}, the PMU-based SE model considered here allows a more explicit interpretation of the resulting convergence condition.

To carry out the analysis, consider a portion of the factor graph in which the pairwise factor node $f_k \in \widetilde{\mathcal{I}}$ is connected to the variable nodes $\mathbf{x}_i, \mathbf{x}_j \in \mathcal{X}$. The corresponding neighboring factor sets are denoted by $\mathcal{F}_i = \mathcal{V}_i \cup \widetilde{\mathcal{I}}_i$ and $\mathcal{F}_j = \mathcal{V}_j \cup \widetilde{\mathcal{I}}_j$, where $\mathcal{V}_i$ and $\mathcal{V}_j$ denote the sets of unary factor nodes connected to $\mathbf{x}_i$ and $\mathbf{x}_j$, respectively, while $\widetilde{\mathcal{I}}_i$ and $\widetilde{\mathcal{I}}_j$ denote the corresponding sets of pairwise factor nodes.

For mathematical consistency of the convergence analysis, we assume that each variable node is initialized with finite uncertainty and that all message precision updates remain well defined throughout the GBP iterations.

\subsection{Convergence of the Message Precision Matrices}
We first analyze the convergence of the GBP message precision matrices. By combining the precision update in \eqref{BP_fv_var} with the precision update in \eqref{BP_vf_var}, a closed recursion is obtained for the precision matrix of the message sent from factor node $f_k \in \widetilde{\mathcal{I}}$ to variable node $\mathbf{x}_i$:
\begin{equation}
	\mathbf{\Lambda}_{f_k \to \mathbf{x}_i}^{(\nu)} =
	\mathbf{H}_{f_k,\mathbf{x}_i}^{T}
	\left[\mathbf{V}_{f_k}^{(\nu)}\right]^{-1}
	\mathbf{H}_{f_k,\mathbf{x}_i},
	\label{eq:fusion_precision_recursion}
\end{equation}
where:
\begin{equation}
	\mathbf{V}_{f_k}^{(\nu)} =
	\mathbf{\Sigma}_{f_k} + \mathbf{H}_{f_k,\mathbf{x}_j}
	\left[\sum_{f_a \in \mathcal{F}_j \setminus \{f_k\}} \mathbf{\Lambda}_{f_a \to \mathbf{x}_j}^{(\nu-1)}\right]^{-1}
	\mathbf{H}_{f_k,\mathbf{x}_j}^{T},
	\label{eq:fusion_innovation_recursion}
\end{equation}
and $\nu = 1, 2, \dots$ denotes the GBP iteration index. An analogous recursion holds for the message sent from $f_k$ to $\mathbf{x}_j$, obtained by exchanging the subscripts $\mathbf{x}_i$ and $\mathbf{x}_j$ and replacing the corresponding neighboring factor sets.

To represent these updates compactly, all precision messages from pairwise factor nodes to variable nodes are stacked into the block-diagonal matrix:
\begin{equation}
	\mathbf{\Lambda}_{\widetilde{\mathcal{I}}\to\mathcal{X}}^{(\nu)}
	=
	\operatorname{Bdiag}
	\left(
	\mathbf{\Lambda}_{f_k \to \mathbf{x}_i}^{(\nu)},
	\mathbf{\Lambda}_{f_k \to \mathbf{x}_j}^{(\nu)}
	\right)_{f_k \in \widetilde{\mathcal{I}}}.
	\label{eq:global_precision_matrix}
\end{equation}
Using \eqref{eq:fusion_precision_recursion} and \eqref{eq:fusion_innovation_recursion}, the stacked precision matrix at iteration $\nu$ is obtained from that at iteration $\nu-1$ as:
\begin{equation}
	\mathbf{\Lambda}_{\widetilde{\mathcal{I}}\to\mathcal{X}}^{(\nu)}
	=
	\Phi\!\left(\mathbf{\Lambda}_{\widetilde{\mathcal{I}}\to\mathcal{X}}^{(\nu-1)}\right),
	\label{eq:global_precision_mapping}
\end{equation}
where $\Phi(\cdot)$ denotes the corresponding update mapping.

Under positive semidefinite initialization of the message precision matrices, Theorems 5 and 6 in \cite{du2018convergence} imply that the mapping in \eqref{eq:global_precision_mapping} admits a unique positive definite fixed point and that the sequence of stacked precision matrices converges to this fixed point:
\begin{equation}
	\lim_{\nu \to \infty}
	\mathbf{\Lambda}_{\widetilde{\mathcal{I}}\to\mathcal{X}}^{(\nu)}
	=
	\mathbf{\Lambda}_{\widetilde{\mathcal{I}}\to\mathcal{X}}^{*}.
	\label{eq:precision_fixed_point}
\end{equation}
Furthermore, the sequence of stacked precision matrices approaches an arbitrarily small neighborhood of the fixed point at a doubly exponential rate \cite{du2018convergence}.

\subsection{Convergence of the Message Mean Vectors}
Having established the convergence of the message precision matrices in \eqref{eq:precision_fixed_point}, we next turn to the convergence of the message mean vectors. In the corresponding update equations, the precision matrices are replaced by their fixed-point values.

At this stage, it is convenient to separate the constant contribution from the iterative one in the message mean update. The constant part collects the contributions of the unary factor nodes and the measurement terms associated with neighboring pairwise factor nodes, while the iterative part captures the dependence on the incoming message mean vectors from the previous GBP iteration.

Accordingly, substituting \eqref{fused_mean_message} into \eqref{BP_vf_mean}, with the precision matrices replaced by their fixed-point values, yields the message sent from variable node $\mathbf{x}_i$ to factor node $f_k \in \widetilde{\mathcal{I}}_i$:
\begin{equation}
	\mathbf{z}_{\mathbf{x}_i \to f_k}^{(\nu)}
	=
	\mathbf{b}_{\mathbf{x}_i \to f_k}
	-
	\sum_{f_a \in \widetilde{\mathcal{I}}_i \setminus \{f_k\}}
	\mathbf{K}_{f_a \to \mathbf{x}_i}
	\mathbf{z}_{\mathbf{x}_j \to f_a}^{(\nu-1)}.
	\label{eq:mean_recursion_affine}
\end{equation}
The constant vector $\mathbf{b}_{\mathbf{x}_i \to f_k} \in \mathbb{R}^{2}$ is given by:
\begin{equation}
\begin{aligned}
	\mathbf{b}_{\mathbf{x}_i \to f_k}
	&=
	\left[\mathbf{\Lambda}_{\mathbf{x}_i \to f_k}^{*}\right]^{-1}
	\Bigg(
	\sum_{f_a \in \mathcal{V}_i}
	\mathbf{\Lambda}_{f_a \to \mathbf{x}_i}\mathbf{z}_{f_a \to \mathbf{x}_i}
	\\
	&\qquad\quad
	+
	\sum_{f_a \in \widetilde{\mathcal{I}}_i \setminus \{f_k\}}
	\mathbf{H}_{f_a,\mathbf{x}_i}^{T}
	\left[\mathbf{V}_{f_a}^{*}\right]^{-1}
	\mathbf{z}_{f_a}
	\Bigg),
\end{aligned}
\label{eq:b_definition}
\end{equation}
where it collects all terms that do not depend on the previous GBP iteration. The matrix $\mathbf{K}_{f_a \to \mathbf{x}_i} \in \mathbb{R}^{2 \times 2}$ is defined as:
\begin{equation}
	\mathbf{K}_{f_a \to \mathbf{x}_i}
	=
	\left[\mathbf{\Lambda}_{\mathbf{x}_i \to f_k}^{*}\right]^{-1}
	\mathbf{H}_{f_a,\mathbf{x}_i}^{T}
	\left[\mathbf{V}_{f_a}^{*}\right]^{-1}
	\mathbf{H}_{f_a,\mathbf{x}_j},
	\label{eq:K_definition}
\end{equation}
and represents the local feedback term through which neighboring pairwise messages influence the current update. Finally, the fixed-point innovation covariance matrix $\mathbf{V}_{f_a}^{*}$ is given by:
\begin{equation}
	\mathbf{V}_{f_a}^{*}
	=
	\mathbf{\Sigma}_{f_a}
	+
	\mathbf{H}_{f_a,\mathbf{x}_j}
	\left[\mathbf{\Lambda}_{\mathbf{x}_j \to f_a}^{*}\right]^{-1}
	\mathbf{H}_{f_a,\mathbf{x}_j}^{T}.
	\label{eq:Vstar_definition}
\end{equation}

Let all message mean vectors from variable nodes to pairwise factor nodes be stacked into the global vector:
\begin{equation}
	\mathbf{z}_{\mathcal{X}\to\widetilde{\mathcal{I}}}^{(\nu)}
	=
	\operatorname{col}\Big(
	\mathbf{z}_{\mathbf{x}_i \to f_k}^{(\nu)},
	\mathbf{z}_{\mathbf{x}_j \to f_k}^{(\nu)}
	\Big)_{f_k \in \widetilde{\mathcal{I}}},
	\label{eq:global_mean_vector}
\end{equation}
and define analogously the constant vector:
\begin{equation}
	\mathbf{b}_{\mathcal{X}\to\widetilde{\mathcal{I}}}
	=
	\operatorname{col}\Big(
	\mathbf{b}_{\mathbf{x}_i \to f_k},
	\mathbf{b}_{\mathbf{x}_j \to f_k}
	\Big)_{f_k \in \widetilde{\mathcal{I}}}.
	\label{eq:global_b_vector}
\end{equation}
Using the same ordering in both stacked vectors, the local recursions in \eqref{eq:mean_recursion_affine} can be written in the compact form:
\begin{equation}
	\mathbf{z}_{\mathcal{X}\to\widetilde{\mathcal{I}}}^{(\nu)}
	=
	\mathbf{b}_{\mathcal{X}\to\widetilde{\mathcal{I}}}
	-
	\mathbf{Q}\,
	\mathbf{z}_{\mathcal{X}\to\widetilde{\mathcal{I}}}^{(\nu-1)},
	\label{eq:global_mean_recursion}
\end{equation}
where $\mathbf{Q}$ is the block matrix formed by the matrices $\mathbf{K}_{f_a \to \mathbf{x}_i}$ in \eqref{eq:K_definition}. The recursion in \eqref{eq:global_mean_recursion} is an affine linear system. Therefore, by Theorem 13 in \cite{du2018convergence}, the message mean vectors converge to a unique fixed point if and only if:
\begin{equation}
	\rho(\mathbf{Q}) < 1,
	\label{eq:rho_Q_condition}
\end{equation}
where $\rho(\mathbf{Q})$ denotes the spectral radius of $\mathbf{Q}$. In that case, the fixed point satisfies:
\begin{equation}
	\mathbf{z}_{\mathcal{X}\to\widetilde{\mathcal{I}}}^{*}
	=
	\left(\mathbf{I} + \mathbf{Q}\right)^{-1}
	\mathbf{b}_{\mathcal{X}\to\widetilde{\mathcal{I}}}.
	\label{eq:mean_fixed_point}
\end{equation}
Finally, once the fixed-point message mean vectors in \eqref{eq:mean_fixed_point} are available, the marginal means are obtained from \eqref{BP_marginal_mean_var}. These marginals yield the local state estimates that solve the PMU-based SE problem in \eqref{likelihood_solution}. Because the estimator in \eqref{likelihood_solution} is equivalent to the centralized WLS estimator in \eqref{wls_solution}, the converged GBP solution recovers the WLS estimate.

\subsection{Convergence and PMU-Based Factor Graph Structure}
The convergence condition in \eqref{eq:rho_Q_condition} shows that the behavior of the message mean vectors is governed by the feedback matrix $\mathbf{Q}$ in \eqref{eq:global_mean_recursion}. This makes it possible to interpret the convergence properties of the proposed PMU-based model directly through the structure of the underlying factor graph.

In the considered PMU-based SE model, the factor graph remains sparse because each bus is physically connected only to a limited number of neighboring buses. The fused representation further simplifies this structure by replacing multiple current phasor measurements associated with the same pair of variable nodes with a single pairwise factor node. As a result, each set $\widetilde{\mathcal{I}}_i$ contains only a small number of pairwise factor nodes, redundant local interactions between the same neighboring variables are removed, and the local recursion in \eqref{eq:mean_recursion_affine} depends on only a limited number of neighboring message mean vectors. Consequently, the corresponding block rows of $\mathbf{Q}$ contain only a small number of nonzero blocks, and the feedback in the global mean recursion remains strongly localized. This graph structure acts in favor of the convergence condition in \eqref{eq:rho_Q_condition}.

The role of unary factor nodes is also favorable. Referring to the local graph structure introduced at the beginning of this section, the sets $\mathcal{V}_i$ and $\mathcal{V}_j$ are either empty or singleton sets, and at least one of them must be nonempty. Otherwise, no PMU would be installed at either end of the branch, and the factor node $f_k$ would not appear in the graph. It follows that the maximum distance from any pairwise factor node to a variable node incident to a unary factor node is one hop. Thus, unary factor nodes remain in immediate proximity to pairwise factor nodes and provide direct local information to neighboring variables.

This structural property is also reflected in the message mean recursion. As seen from \eqref{eq:mean_recursion_affine}, unary contributions enter only through the constant term $\mathbf{b}_{\mathbf{x}_i \to f_k}$, whereas the iterative part is governed by the matrices $\mathbf{K}_{f_a \to \mathbf{x}_i}$ in \eqref{eq:K_definition}. Unary factor nodes therefore do not introduce additional feedback paths in the mean recursion. Instead, they anchor the local message updates through direct local information, while the recursive feedback remains confined to the pairwise part of the graph. The same effect is also reflected in the term $\mathbf{\Lambda}_{\mathbf{x}_i \to f_k}^{*}$ appearing in \eqref{eq:b_definition} and \eqref{eq:K_definition}. Since unary factor nodes contribute through the precision update in \eqref{BP_vf_var}, stronger local unary information increases the corresponding precision term. As a result, the contribution of neighboring pairwise interactions to the matrices $\mathbf{K}_{f_a \to \mathbf{x}_i}$ is effectively moderated, which acts in favor of reducing the feedback strength in the mean recursion. In practical PMU implementations, this effect may be further reinforced by the fact that voltage phasor measurements are often modeled as more accurate than current phasor measurements, since current measurements can be affected by additional uncertainty introduced by current-channel instrumentation, particularly current transformers \cite{davanzo2020ct}.

Taken together, the sparse PMU-based graph structure, the fusion of redundant pairwise interactions, and the local anchoring introduced by unary factor nodes all act in favor of the convergence condition in \eqref{eq:rho_Q_condition}. Although these structural properties do not by themselves constitute a general proof that $\rho(\mathbf{Q}) < 1$ always holds, they provide a clear explanation for the robust convergence observed in the proposed fusion-based formulation. In particular, the numerical results show that the corresponding mean-message recursion converges consistently across the considered power systems and a broad range of PMU placements and redundancy levels. These findings further reinforce the practical relevance of the proposed fusion-based formulation for PMU-based SE.

\section{Numerical Results}
The proposed GBP-based algorithms are evaluated in terms of convergence, accuracy, scalability, and performance under asynchronous measurement arrivals. In all simulation scenarios, the measurement set is generated using an open-source software package \cite{cosovic2025juliagrid}. For each considered power system test case, AC power flow is first solved to obtain the exact system state, except in the last subsection, where AC optimal power flow analysis is used instead. The resulting bus voltages and branch currents are then corrupted by additive white Gaussian noise with specified magnitude and angle variances for voltage and current measurements in order to generate synthetic phasor measurements. For each simulation, PMUs are initially placed using the optimal placement algorithm in \cite{gou2008optimal}, thereby ensuring full system observability. In selected scenarios, additional PMUs are placed randomly on top of the optimal configuration in order to obtain statistically significant results and assess the impact of measurement redundancy and varying measurement configurations on the proposed method.

\subsection{Convergence Rate and Accuracy of the GBP Algorithms}
In the first scenario, we evaluate the convergence rate and accuracy of the scalar, multivariate, and fusion-based GBP algorithms relative to the centralized WLS solution under the assumption that measurement covariances are neglected. The analysis is conducted on a 1354-bus power system \cite{fliscounakis2013contingency}, where PMUs are optimally placed to ensure system observability, resulting in 397 PMUs and 2397 phasor measurements. Each voltage magnitude and angle measurement is assigned a variance of $\sigma^2 = 10^{-8}$, while each current magnitude and angle measurement is assigned a variance of $\sigma^2 = 10^{-6}$. The main characteristics of the corresponding factor graphs are summarized in Table~\ref{factor_graph_1354}, showing that the fused formulation is considerably less complex than the other approaches.
\begin{table}[ht]
	\centering
	\vspace{2pt} 
	\caption{Comparison of scalar, multivariate, and fusion factor graphs}
	\label{factor_graph_1354}
	\footnotesize
	\setlength{\tabcolsep}{8pt}
	\begin{tabular}{l r r r}
		\midrule
		& Variable Nodes & Factor Nodes & Pairwise Edges \\
		\midrule
		Scalar        & 2708 & 4794 & 15966 \\
		Multivariate  & 1354 & 2397 &  3992 \\
		Fusion  	      & 1354 & 1849 &  2898 \\
		\midrule
	\end{tabular}
	\vspace{2pt} 
\end{table}

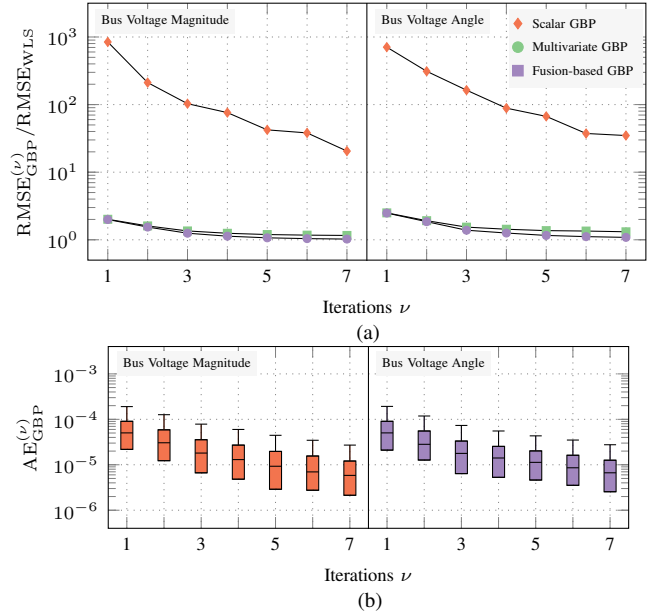
\begin{figure}[ht]
	\centering \captionsetup[subfigure]{oneside,margin={1.3cm,0cm}}
	\begin{tabular}{@{\hspace{-0.2cm}}c@{}} 
	\subfloat[]{\label{plot1a} \centering	
    \begin{tikzpicture}
    
	\begin{axis}[
		width = 9cm, height = 5.0cm, ymode = log,
		grid style = {line width=.3pt, draw=gray},
		x tick label style = {/pgf/number format/.cd, set thousands separator={},fixed},
		ylabel style = {yshift=-0.1cm},
		xlabel = {Iterations $\nu$},
		ylabel = {$\mathrm{RMSE}^{(\nu)}_{\mathrm{GBP}} / \mathrm{RMSE}_{\mathrm{WLS}}$},
		label style = {font=\scriptsize},
		grid = major,
		ymin = 0.5, ymax = 10^3.5, 
		xmin = 0.5, xmax = 14.5,
		xtick = {1,2,3,4,5,6,7, 8,9,10,11,12,13,14},
		xticklabels = {1,,3,,5,,7, 1,,3,,5,,7},
		ytick={10^0, 10^1, 10^2, 10^3},
		tick label style = {font=\scriptsize}, 
		legend style={draw=none, fill=gray!7!white, legend cell align=left, font=\tiny, at={(0.995, 0.978)}, anchor=north east, 
    		/tikz/column 1/.append style={column sep=0.05cm}, inner sep=1.5pt,
		/tikz/column 2/.style={yshift=-0.02cm}}, 
		legend columns = 1
		]
                
		\addlegendimage{mark=diamond*, only marks, mark options={wind}}
		\addlegendentry{Scalar GBP}; 
		\addlegendimage{mark=otimes*, only marks, mark options={gas}}
		\addlegendentry{Multivariate GBP}; 
		\addlegendimage{mark=square*, only marks, mark options={solar}}
		\addlegendentry{Fusion-based GBP}; 

		\addplot[mark=diamond*, mark size=1.7pt, mark options={wind}, black] 
		table [x={iter}, y={scalar}] {./plot/plot1/rmseMagnitude.txt}; 
			
		\addplot[mark=square*, mark size=1.5pt, mark options={gas}, black] 
		table [x={iter}, y={vector}] {./plot/plot1/rmseMagnitude.txt}; 
  	            
		\addplot[mark=otimes*, mark size=1.5pt, mark options={solar}, black]
		table [x={iter}, y={fusion}] {./plot/plot1/rmseMagnitude.txt}; 

		\addplot[mark=diamond*, mark size=1.7pt, mark options={wind}, black] 
		table [x={iter}, y={scalar}] {./plot/plot1/rmseAngle.txt}; 
	        
		\addplot[mark=square*, mark size=1.5pt, mark options={gas}, black] 
		table [x={iter}, y={vector}] {./plot/plot1/rmseAngle.txt}; 
  	        
  	  	\addplot[mark=otimes*, mark size=1.5pt, mark options={solar}, black]
		table [x={iter}, y={fusion}] {./plot/plot1/rmseAngle.txt}; 
  	            
		\draw [thin] (axis cs:7.5,0.00001) -- (axis cs:7.5, 10000);  	            
		\node[fill=gray!7!white, inner sep=2.3pt, anchor=south east] at (axis cs: 4.2, 10^3.05) {\tiny{$\text{Bus Voltage Magnitude}$}};
		\node[fill=gray!7!white, inner sep=2.3pt, anchor=south east] at (axis cs: 10.6, 10^3.05) {\tiny{$\text{Bus Voltage Angle}$}};
	\end{axis} 
		
	\end{tikzpicture}}
	\end{tabular} 
	\captionsetup[subfigure]{oneside,margin={1.5cm,0cm}}
	
	\begin{tabular}{@{\hspace{-0.35cm}}c@{\hspace{0cm}}} \subfloat[]{\label{plot1b} \centering
	\begin{tikzpicture}
	
	\begin{axis}[
		box plot width = 0.8mm, ymode = log, width = 8.5cm, height = 4.0cm,
		grid style = {line width=.3pt, draw=gray},
		xlabel style = {align=center},
		xlabel = {Iterations $\nu$},
		ylabel = {$\mathrm{AE}^{(\nu)}_{\mathrm{GBP}}$},
        	ylabel style = {yshift=-0.1cm},
  	    grid = major,   		
  	    xmin = 0.5, xmax = 14.5,
		ymin = 4e-7, ymax = 4e-3, 	
		ytick = {1e-6, 1e-5, 1e-4, 1e-3},
  	    xtick = {1,2,3,4,5,6,7, 8,9,10,11,12,13,14},
  	    xticklabels = {1,,3,,5,,7, 1,,3,,5,,7},
  	    tick label style = {font=\scriptsize}, label style={font=\scriptsize},
  	  	]     
	
		\boxplot[
			forget plot, fill=wind, box plot whisker bottom index=2,
			box plot whisker top index=5, box plot box bottom index=2,
            	box plot box top index=4, box plot median index=3
    		] 
       	{./plot/plot1/boxMagnitude.txt};
            
		\draw [thin] (axis cs:7.5, 0.1) -- (axis cs:7.5, 1e-7);
                
		\boxplot[
			forget plot, fill=solar, box plot whisker bottom index=2,
			box plot whisker top index=5, box plot box bottom index=2,
            	box plot box top index=4, box plot median index=3
    		] 
        {./plot/plot1/boxAngle.txt};
                
		\node[fill=gray!7!white, inner sep=2.3pt, anchor=south east] at (axis cs: 4.5, 0.9e-3) {\tiny{$\text{Bus Voltage Magnitude}$}};
		\node[fill=gray!7!white, inner sep=2.3pt, anchor=south east] at (axis cs: 10.8, 0.9e-3) {\tiny{$\text{Bus Voltage Angle}$}};
	\end{axis}
		
	\end{tikzpicture}}
	\end{tabular}
	\caption{
		Normalized RMSE of the scalar, multivariate, and fusion-based GBP algorithms over iterations $\nu$, relative to the 
		WLS RMSE for bus voltage magnitudes and angles (subfigure a), and the component-wise absolute error of the fusion-based GBP 
		algorithm with respect to the WLS estimates (subfigure b). For clarity, the lower whiskers are truncated, as they correspond to 
		very small errors, while the focus is on the upper range of the error distribution.
	}
	\label{plot1}
\end{figure} 

The convergence behavior of the scalar, multivariate, and fusion-based GBP formulations is illustrated in \figurename~\ref{plot1a} through the ratio of the root mean square errors (RMSE), defined as $\mathrm{RMSE}^{(\nu)}_{\mathrm{GBP}} / \mathrm{RMSE}_{\mathrm{WLS}}$, for both voltage magnitudes and angles. Here, $\mathrm{RMSE}^{(\nu)}_{\mathrm{GBP}}$ is computed from the GBP estimates at iteration $\nu$, while $\mathrm{RMSE}_{\mathrm{WLS}}$ is computed from the centralized WLS solution in \eqref{wls_solution}. Both metrics are evaluated with respect to the exact solution obtained from power flow analysis. As the ratio approaches one, the GBP estimates converge to the WLS solution. The results show that the multivariate and fusion-based GBP formulations converge significantly faster than the scalar formulation, with the fused formulation exhibiting the fastest convergence. Since the fusion-based GBP formulation exhibits the best convergence rate, the following analyses focus on this formulation.

A more detailed view of the fusion-based GBP performance is provided in \figurename~\ref{plot1b} through the component-wise absolute error (AE) relative to the WLS solution. The AE is computed separately for voltage magnitude and angle estimates at each GBP iteration $\nu$, yielding the vectors $\mathrm{AE}^{(\nu)}_{\mathrm{GBP}}$, which are visualized using box plots to illustrate the error distribution. As shown, after just one iteration, the fusion-based GBP algorithm achieves errors below $10^{-4}$ for $75\%$ of the estimated voltage magnitudes and angles, while most of the remaining estimates exhibit only slightly larger errors. After three iterations, $98\%$ of the estimates fall below the $10^{-4}$ threshold. For this high-voltage network, operating at line-to-line voltages of $380\,\mathrm{kV}$ and $220\,\mathrm{kV}$, this corresponds to an approximate line-to-neutral voltage error between $13\,\mathrm{V}$ and $22\,\mathrm{V}$, while the corresponding angle error is approximately $0.0057^\circ$. These results indicate that the fusion-based GBP algorithm converges rapidly and is well suited for near real-time SE.

\subsection{Convergence of the Fusion-Based GBP Algorithm}
\begin{figure}[ht]
	\centering \captionsetup[subfigure]{oneside,margin={1.4cm,0cm}}
	\begin{tabular}{@{\hspace{-0.35cm}}c@{\hspace{0cm}}} \subfloat[]{\label{plot2a} \centering	
    \begin{tikzpicture}
        
  	    \begin{axis} [
  	    		box plot width = 0.8mm, width = 8.5cm, height = 4.0cm, ymode = log,  
			log plot exponent style/.style={/pgf/number format/fixed, /pgf/number format/precision=0},
			grid style = {line width=.3pt, draw=gray},
			y tick label style = {/pgf/number format/.cd,fixed, fixed zerofill, precision=2, /tikz/.cd},
	        	xlabel style = {align=center},
    	        	xlabel = {Power System},
        	    	ylabel = {$1 - \rho(\mathbf{Q})$},
        	    	ylabel style ={yshift=-0.1cm},
  	        grid = major,   		
  	       	xmin = 1, xmax = 13,
			ymin = 2e-6, ymax = 5e-1,
			ytick = {1e-5, 1e-4,1e-3,1e-2,1e-1},
  	        xtick = {2,4,6,8,10,12},
  	        xticklabels = {60-bus,89-bus,200-bus,300-bus,500-bus,1354-bus},
  	        tick label style = {font=\scriptsize}, 
  	        label style = {font=\scriptsize},
  	        ]     
	
	       	\boxplot [
	       		forget plot, fill=wind, 
           		box plot whisker top index=5, box plot box bottom index=2,
				box plot box top index=4, box plot median index=3, box plot whisker bottom index=1
          	] 
            	{./plot/plot2/optimal_placement2.txt};
		\end{axis}
	        
	\end{tikzpicture}}
	\end{tabular}
	\captionsetup[subfigure]{oneside,margin={1.5cm,0cm}}
	
	\begin{tabular}{@{\hspace{-0.35cm}}c@{\hspace{0cm}}} \subfloat[]{\label{plot2b} \centering	
   	\begin{tikzpicture}
   	
  	    \begin{axis} [
  	    		box plot width = 0.8mm, width = 8.5cm, height = 4.0cm, ymode = log,  
			log plot exponent style/.style={/pgf/number format/fixed, /pgf/number format/precision=0},
			grid style = {line width=.3pt, draw=gray},
			y tick label style = {/pgf/number format/.cd,fixed, fixed zerofill, precision=2, /tikz/.cd},
	        	xlabel style = {align=center},
    	        	xlabel = {Power System},
        	    	ylabel = {$1 - \rho(\mathbf{Q})$},
        	    	ylabel style ={yshift=-0.1cm},
  	        grid = major,   		
  	       	xmin = 1, xmax = 13,
			ymin = 3e-4, ymax = 4,
			ytick = {1e-3,1e-2,1e-1, 1},
  	        xtick = {2,4,6,8,10,12},
  	        xticklabels = {60-bus,89-bus,200-bus,300-bus,500-bus,1354-bus},
  	        tick label style = {font=\scriptsize}, 
  	        label style = {font=\scriptsize},
  	        ]     
	
	       	\boxplot [
	       		forget plot, fill=wind, 
           		box plot whisker top index=5, box plot box bottom index=2,
				box plot box top index=4, box plot median index=3, box plot whisker bottom index=1
          	] 
            	{./plot/plot2/redundancy2.txt};
		\end{axis}
		
	\end{tikzpicture}}
	\end{tabular}
	\caption{
		The quantity $1 - \rho(\mathbf{Q})$ for the fusion-based GBP algorithm for power systems with 60, 89, 200, 300, 500, and 1354 buses,
		obtained for optimal PMU placement with varying voltage measurement variances and fixed current 
		measurement variances (subfigure a), and for scenarios with additional PMUs randomly added to the optimal placement under 
		fixed voltage and current measurement variances (subfigure b).
	}
	\label{plot2}
\end{figure}
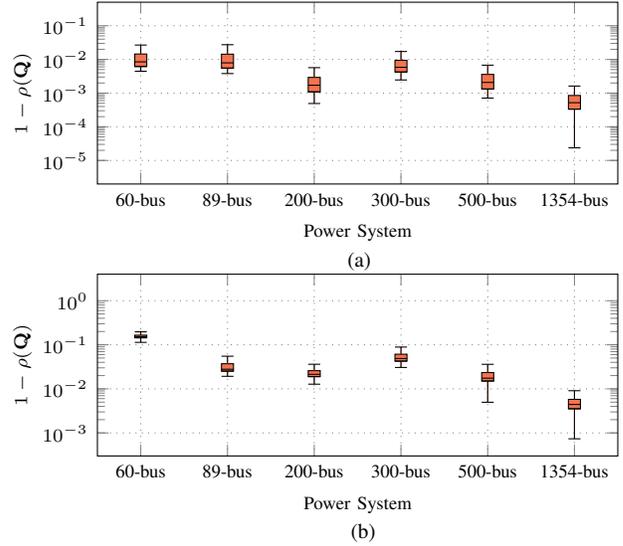

The convergence of the fusion-based GBP algorithm is further examined through the spectral radius of the matrix $\mathbf{Q}$ in \eqref{eq:global_mean_recursion} for the 60-bus \cite{capitanescu2020}, 89-bus \cite{fliscounakis2013contingency}, 200-bus \cite{birchfield2016grid}, IEEE 300-bus, 500-bus \cite{birchfield2016grid}, and 1354-bus power systems. The results are presented in \figurename~\ref{plot2} using box plots, where each box plot represents the distribution over 1000 simulations. To better highlight differences across systems, $1 - \rho(\mathbf{Q})$ is plotted instead of $\rho(\mathbf{Q})$, so positive values indicate that the convergence condition is satisfied.

The results for optimal PMU placement are shown in \figurename~\ref{plot2a}. In each simulation, the voltage magnitude and angle variances are randomly selected from $\sigma^2 \in [10^{-8}, 10^{-6}]$, while the current magnitude and angle variances are fixed at $\sigma^2 = 10^{-6}$. This setting examines convergence under different levels of unary factor node precision, including the case where unary and pairwise factor nodes have the same precision. All plotted values of $1 - \rho(\mathbf{Q})$ remain positive, and no outliers extend below zero.

To further examine the effect of measurement redundancy, \figurename~\ref{plot2b} considers a scenario in which, starting from the optimal PMU placement, additional PMUs are randomly added at buses without PMUs. For each simulation, a probability $p \in [0.2,0.8]$ is randomly generated, and a PMU is independently installed at each such bus with probability $p$. Here, the voltage magnitude and angle variances are fixed at $\sigma^2 = 10^{-8}$, while the current magnitude and angle variances are fixed at $\sigma^2 = 10^{-6}$. This enables the convergence analysis under varying measurement configurations, redundancy levels, and system sizes. The plotted values remain positive in all simulations, with no outliers extending below zero, confirming robust convergence across all considered power systems.

Taken together, these results provide strong empirical evidence that the fusion-based GBP algorithm exhibits robust convergence in PMU-based SE. Across all considered power systems, measurement configurations, redundancy levels, and measurement variance settings, the quantity $1 - \rho(\mathbf{Q})$ remains positive, which means that the spectral radius stays below one and the mean-message recursion converges asymptotically to its fixed point. At the same time, a spectral radius close to one does not necessarily imply slow convergence to a practically relevant neighborhood of the fixed point within a finite number of iterations, since it characterizes asymptotic behavior rather than finite-iteration performance.

\subsection{Scalability of the Fusion-Based GBP Algorithm}
Scalability is evaluated in terms of convergence rate and accuracy on the 1354-bus and 13659-bus power systems \cite{fliscounakis2013contingency}, while accounting for measurement covariances in both cases. The voltage magnitude and angle variances are fixed at $\sigma^2 = 10^{-8}$, while the current magnitude and angle variances are fixed at $\sigma^2 = 10^{-6}$. Starting from the optimal PMU placement, additional PMUs are randomly added at buses without PMUs. For each of the 1000 simulation runs, a probability $p \in [0.2, 0.8]$ is randomly generated, and a PMU is independently installed at each such bus with probability $p$. 

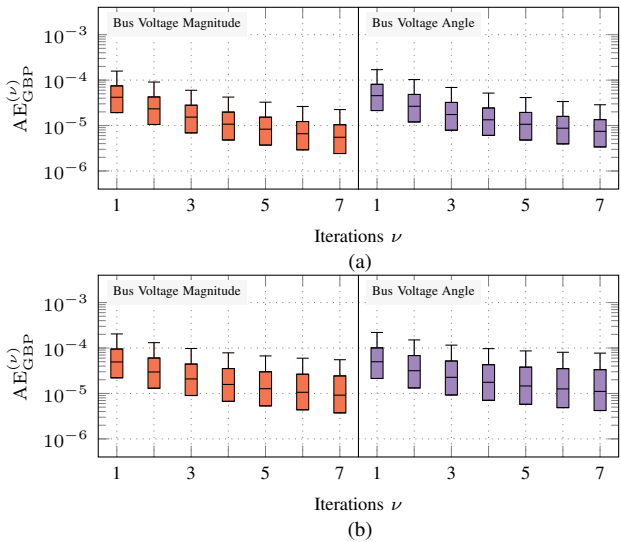
\begin{figure}[ht]
	\centering \captionsetup[subfigure]{oneside,margin={1.5cm,0cm}}
	\begin{tabular}{@{\hspace{-0.35cm}}c@{\hspace{0cm}}} \subfloat[]{\label{plot3a} \centering	
    \begin{tikzpicture}
        
 	\begin{axis} [
  	  	box plot width = 0.8mm, ymode = log, width = 8.5cm, height = 4.0cm,
		grid style = {line width=.3pt, draw=gray},
	    	xlabel style = {align=center},
    	    xlabel = {Iterations $\nu$},
        ylabel = {$\mathrm{AE}^{(\nu)}_{\mathrm{GBP}}$},
        	ylabel style ={yshift=-0.1cm},
  	    grid = major,   		
  	    xmin = 0.5, xmax = 14.5,
		ymin = 4e-7, ymax = 4e-3, 	
		ytick = {1e-6, 1e-5, 1e-4, 1e-3}, 
  	    xtick = {1,2,3,4,5,6,7, 8,9,10,11,12,13,14},
  	    xticklabels = {1,,3,,5,,7, 1,,3,,5,,7},
  	   	tick label style = {font=\scriptsize}, 
  	    label style = {font=\scriptsize},
  	  	]     
	
	    \boxplot [
	       	forget plot, fill=wind, box plot whisker bottom index=2,
           	box plot whisker top index=5, box plot box bottom index=2,
			box plot box top index=4, box plot median index=3
       	] 
      	{./plot/plot3/box_case1354_p02_08_magnitude.txt};
            	
      	\draw [thin] (axis cs:7.5, 0.1) -- (axis cs:7.5, 1e-7);
                
 		\boxplot [
			forget plot, fill=solar, box plot whisker bottom index=2,
			box plot whisker top index=5, box plot box bottom index=2,
			box plot box top index=4, box plot median index=3
        ] 
        	{./plot/plot3/box_case1354_p02_08_angle.txt};
                
      	\node[fill=gray!7!white, inner sep=2.3pt, anchor=south east] at (axis cs: 4.5, 0.9e-3) {\tiny{$\text{Bus Voltage Magnitude}$}};
  	   	\node[fill=gray!7!white, inner sep=2.3pt, anchor=south east] at (axis cs: 10.8, 0.9e-3) {\tiny{$\text{Bus Voltage Angle}$}};
  	\end{axis}
	        
	\end{tikzpicture}}
	\end{tabular}
	\captionsetup[subfigure]{oneside,margin={1.5cm,0cm}}
	
	\begin{tabular}{@{\hspace{-0.35cm}}c@{\hspace{0cm}}} \subfloat[]{\label{plot3b} \centering	
   	\begin{tikzpicture}
   	
	\begin{axis} [
	  	box plot width = 0.8mm, ymode = log, width = 8.5cm, height = 4.0cm,
	   	grid style = {line width=.3pt, draw=gray},
		xlabel style = {align=center},
		xlabel = {Iterations $\nu$},
        	ylabel = {$\mathrm{AE}^{(\nu)}_{\mathrm{GBP}}$},
        	ylabel style = {yshift=-0.1cm},
  	    grid = major,   		
		xmin = 0.5, xmax = 14.5,
		ymin = 4e-7, ymax = 4e-3, 	
		ytick = {1e-6, 1e-5, 1e-4, 1e-3},
		xtick = {1,2,3,4,5,6,7, 8,9,10,11,12,13,14},
		xticklabels = {1,,3,,5,,7, 1,,3,,5,,7},
		tick label style = {font=\scriptsize}, label style={font=\scriptsize},
		]     
	
		\boxplot[
			forget plot, fill=wind, box plot whisker bottom index=2,
			box plot whisker top index=5, box plot box bottom index=2,
			box plot box top index=4, box plot median index=3
		] 
		{./plot/plot3/box_case13659_p02_08_magnitude.txt};
                
		\draw [thin] (axis cs:7.5,0.1) -- (axis cs:7.5, 1e-7);
                
		\boxplot[
			forget plot, fill=solar, box plot whisker bottom index=2,
			box plot whisker top index=5, box plot box bottom index=2,
			box plot box top index=4, box plot median index=3
		] 
		{./plot/plot3/box_case13659_p02_08_angle.txt};
                
		\node[fill=gray!7!white, inner sep=2.3pt, anchor=south east] at (axis cs: 4.5, 0.9e-3) {\tiny{$\text{Bus Voltage Magnitude}$}};
		\node[fill=gray!7!white, inner sep=2.3pt, anchor=south east] at (axis cs: 10.8, 0.9e-3) {\tiny{$\text{Bus Voltage Angle}$}};
	\end{axis}
		
	\end{tikzpicture}}
	\end{tabular}
	\caption{
		Component-wise absolute error of the fusion-based GBP algorithm with respect to the WLS estimates for the 
		1354-bus system (subfigure a) and the 13659-bus system (subfigure b). For clarity, the lower whiskers are truncated, 
		as they correspond to very small errors, while the focus is on the upper range of the error distribution.
	}
	\label{plot3}
\end{figure}

For the 1354-bus power system, the component-wise absolute error $\mathrm{AE}^{(\nu)}_{\mathrm{GBP}}$ across GBP iterations $\nu$ is shown in \figurename~\ref{plot3a} using box plots cumulatively stacked over 1000 simulations. Compared with \figurename~\ref{plot1b}, no degradation in convergence or accuracy is observed when measurement covariances are taken into account and redundancy is increased. The corresponding results for the 13659-bus power system, shown in \figurename~\ref{plot3b}, further indicate that increasing the system size does not degrade convergence or accuracy. These results confirm that the proposed fusion-based GBP algorithm remains accurate and robust as system size grows.

\subsection{Fusion-Based GBP with Asynchronous PMU Updates}
This subsection evaluates the fusion-based GBP algorithm under asynchronous phasor measurement updates on the 13659-bus power system. The system evolves through a sequence of quasi-stationary operating conditions occurring every $20\,\mathrm{ms}$. PMUs are initially placed according to the optimal placement configuration, and a slight degree of redundancy is introduced by randomly installing additional PMUs at buses without PMUs with probability $p = 0.1$, while the measurement variances remain the same as in the previous subsection. The fusion-based GBP algorithm is then executed continuously. At each iteration, only $60\%$ of the phasor measurements are randomly updated, whereas the remaining measurements retain their previously received values. For measurements that are not updated, the associated variances are increased by a factor of $10^2$ at each iteration to model information aging until a new measurement becomes available. Updated voltage phasor measurements at unary factor nodes are also immediately propagated to all connected pairwise factor nodes, since they better reflect the current system state.

In this setting, the ability of the fusion-based GBP algorithm to track successive operating conditions depends both on the duration of the GBP iterations and on the informativeness of the newly received measurement set at each update instant. For the considered power system, one GBP iteration in canonical form,\footnote{The GBP implementation is available at \url{https://github.com/mcosovic/pmu-gbp-state-estimation}.} as described in Appendix A, takes on average $2.2\,\mathrm{ms}$, including the computation of all messages exchanged in both directions between pairwise factor nodes and variable nodes along 30960 pairwise edges, as well as the marginal updates, whereas obtaining the centralized WLS solution requires $38.5\,\mathrm{ms}$.\footnote{The evaluations are performed on a system running 64-bit Ubuntu 20.04.6 LTS, equipped with an AMD Ryzen 9 5950X processor (16 cores, 32 threads), 32 GB of RAM, and a 1 TB SSD, using Julia 1.10.11 (LTS).} This difference in computational times allows the continuously running GBP algorithm to track changes in the system state at a finer temporal resolution.

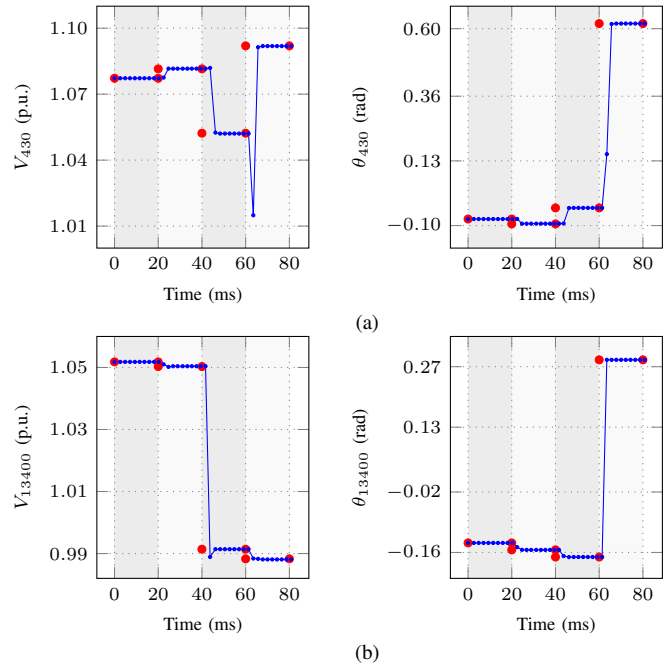
\begin{figure}[ht]
	\centering \captionsetup[subfigure]{oneside,margin={1.0cm,0cm}}
	\begin{tabular}{@{\hspace{-0.35cm}}c@{\hspace{0cm}}} \subfloat[]{\label{plot4a} \centering	
    \begin{tikzpicture}
  		
  	\begin{axis}[
  		width = 4.4cm, height = 4.8cm, 
   		y tick label style = {/pgf/number format/.cd,fixed, fixed zerofill, precision=2, /tikz/.cd},
   		x tick label style = {/pgf/number format/.cd, set thousands separator={},fixed},
   		xlabel = {Time (ms)},
   		ylabel = {$V_{430}$ (p.u.)},
   		label style = {font = \scriptsize},   	
   		grid = major,
		ytick = {1.01, 1.04, 1.07, 1.10},
		ymin = 1.00, ymax = 1.11,
   		tick label style = {font = \scriptsize}, 
   		set layers
   		]
   
   		\begin{pgfonlayer}{axis background}
			\path[fill=gray!15] (axis cs:0,  1.00) rectangle (axis cs:20, 1.11);
			\path[fill=gray!5]  (axis cs:20, 1.00) rectangle (axis cs:40, 1.11);
			\path[fill=gray!15] (axis cs:40, 1.00) rectangle (axis cs:60, 1.11);
			\path[fill=gray!5]  (axis cs:60, 1.00) rectangle (axis cs:80, 1.11);
   		\end{pgfonlayer}
   			
   		\addplot [red, mark=*, only marks, mark indices={1, 2, 3, 4, 5, 6, 7, 8}, mark options={solid}, mark size=1.5pt] 
   		coordinates {
   			(0,  1.077230) (20, 1.077230) 		
   			(20, 1.081492) (40, 1.081492) 
   			(40, 1.052248) (60, 1.052248) 
   			(60, 1.091964) (80, 1.091964)
   		};
   	
   		\addplot [blue,  mark=*, mark size=0.6pt] 
   		coordinates {
   			(0, 1.077230) (2.63, 1.077230) (4.85, 1.077230) (7.08, 1.077230) (9.30, 1.077230) (11.50, 1.077230) 
   			(13.69, 1.077230) (15.89, 1.077230) (18.08, 1.077230) (20.28, 1.077230)
   			(22.47, 1.077536) (24.69, 1.081601) (26.79, 1.081600) (28.86, 1.081600)
			(31.19, 1.081600) (33.33, 1.081600) (35.36, 1.081600) (37.48, 1.081600) (39.53, 1.081600) (41.64, 1.081600)    		
   			(43.72, 1.081911) (46.17, 1.052512) (48.33, 1.052072) (50.50, 1.052072) (52.67, 1.052072)
   			(54.83, 1.052072) (57.00, 1.052072) (59.17, 1.052072) (61.33, 1.052072)
   			(63.50, 1.014920) (65.67, 1.091341) (67.83, 1.091851)
   			(70.00, 1.091851) (72.16, 1.091851) (74.33, 1.091851) (76.50, 1.091851) (78.66, 1.091851) (80.83, 1.091851)
   		};
   	\end{axis}
 
    	\begin{axis}[
    		width = 4.4cm,height = 4.8cm, at = {(4.7cm,0cm)}, 
   		y tick label style = {/pgf/number format/.cd,fixed, fixed zerofill, precision=2, /tikz/.cd},
   		x tick label style = {/pgf/number format/.cd, set thousands separator={},fixed},
   		xlabel = {Time (ms)},
   		ylabel = {$\theta_{430}$ (rad)},
   		label style = {font=\scriptsize},   	
   		grid = major,
		ytick = {-0.1, 0.13, 0.36, 0.6},
		ymin = -0.18, ymax = 0.68,
   		tick label style={font=\scriptsize},
   		set layers
   		]
   		
   		\begin{pgfonlayer}{axis background}
			\path[fill=gray!15] (axis cs:0,  -0.18) rectangle (axis cs:20, 0.68);
			\path[fill=gray!5]  (axis cs:20, -0.18) rectangle (axis cs:40, 0.68);
			\path[fill=gray!15] (axis cs:40, -0.18) rectangle (axis cs:60, 0.68);
			\path[fill=gray!5]  (axis cs:60, -0.18) rectangle (axis cs:80, 0.68);   		
   		\end{pgfonlayer}
   		
   		\addplot [red, mark=*, only marks, mark indices={1, 2, 3, 4, 5, 6, 7, 8}, mark options={solid}, mark size=1.5pt] 
   		coordinates {
   			(0,  -0.076687) (20, -0.076687) 		
   			(20, -0.094048) (40, -0.094048) 
   			(40, -0.037095) (60, -0.037095) 
   			(60,  0.618220) (80,  0.618220)
   		};
   	
   		\addplot [blue,  mark=*, mark size=0.6pt] 
   		coordinates {
   			(0, -0.076695) (2.63, -0.076695) (4.85, -0.076695) (7.08, -0.076695) (9.30, -0.076695) (11.50, -0.076695)
   			(13.69, -0.076695) (15.89, -0.076695) (18.08, -0.076695) (20.28, -0.076695)
   			(22.47, -0.076454) (24.69, -0.093101) (26.79, -0.093100) (28.86, -0.093100)
			(31.19, -0.093100) (33.33, -0.093100) (35.36, -0.093100) (37.48, -0.093100) (39.53, -0.093100) (41.64, -0.093100)
   			(43.72, -0.092497) (46.17, -0.037561) (48.33, -0.037086) (50.50, -0.037086) 
   			(52.67, -0.037086) (54.83, -0.037086) (57.00, -0.037086) (59.17, -0.037086) (61.33, -0.037086)
   			(63.50, 0.154012) (65.67, 0.616711) (67.83, 0.618309)
   			(70.00, 0.618309) (72.16, 0.618309) (74.33, 0.618309) (76.50, 0.618309) (78.66, 0.618309) (80.83, 0.618309)   		
   		};
   	\end{axis}
   	
	\end{tikzpicture}}
	\end{tabular}
	\captionsetup[subfigure]{oneside,margin={1.0cm,0cm}}
	
	\begin{tabular}{@{\hspace{-0.35cm}}c@{\hspace{0cm}}} \subfloat[]{\label{plot4b} \centering	
    \begin{tikzpicture}
  	
  	\begin{axis}[
  		width = 4.4cm, height = 4.8cm, 
   		y tick label style = {/pgf/number format/.cd, fixed, fixed zerofill, precision=2, /tikz/.cd},
   		x tick label style = {/pgf/number format/.cd, set thousands separator={},fixed},
   		xlabel = {Time (ms)},
   		ylabel = {$V_{13400}$ (p.u.)},
   		label style = {font=\scriptsize},   	
   		grid = major,
		ytick = {0.99, 1.01, 1.03, 1.05},
		ymin = 0.982, ymax = 1.06,
   		tick label style = {font=\scriptsize},
   		set layers
   		]
   		
   		\begin{pgfonlayer}{axis background}
			\path[fill=gray!15] (axis cs:0,  0.982) rectangle (axis cs:20, 1.06);
			\path[fill=gray!5]  (axis cs:20, 0.982) rectangle (axis cs:40, 1.06);
			\path[fill=gray!15] (axis cs:40, 0.982) rectangle (axis cs:60, 1.06);
			\path[fill=gray!5]  (axis cs:60, 0.982) rectangle (axis cs:80, 1.06);     		
   		\end{pgfonlayer}
   		
   		\addplot [red, mark=*, only marks, mark indices={1, 2, 3, 4, 5, 6, 7, 8}, mark options={solid}, mark size=1.5pt] 
   		coordinates {
   			(0,  1.0518) (20, 1.0518) 		
   			(20, 1.0503) (40, 1.0503) 
   			(40, 0.9914) (60, 0.9914) 
   			(60, 0.9883) (80, 0.9883)
   		};
   	
   		\addplot [blue,  mark=*, mark size=0.6pt] 
   		coordinates {
   			(0, 1.0518) (2.63, 1.0518) (4.85, 1.0518) (7.08, 1.0518) (9.30, 1.0518) (11.50, 1.0518)
   			(13.69, 1.0518) (15.89, 1.0518) (18.08, 1.0518) (20.28, 1.0518)
   			(22.47, 1.0510) (24.69, 1.0502) (26.79, 1.0504) (28.86, 1.0504)
			(31.19, 1.0504) (33.33, 1.0504) (35.36, 1.0504) (37.48, 1.0504) (39.53, 1.0504) (41.64, 1.0504)
   			(43.72, 0.9889) (46.17, 0.9914) (48.33, 0.9914) (50.50,0.9914) 
   			(52.67, 0.9914) (54.83, 0.9914) (57.00, 0.9914) (59.17, 0.9914) (61.33, 0.9914)
   			(63.50, 0.9884) (65.67, 0.9883) (67.83, 0.9881)
   			(70.00, 0.9881) (72.16, 0.9881) (74.33, 0.9881) (76.50, 0.9881) (78.66, 0.9881) (80.83, 0.9881)    		
   		};
   	\end{axis}
  
    	\begin{axis}[
    		width = 4.4cm,height = 4.8cm, at = {(4.7cm,0cm)}, 
   		y tick label style = {/pgf/number format/.cd,fixed, fixed zerofill, precision=2, /tikz/.cd},
   		x tick label style = {/pgf/number format/.cd, set thousands separator={},fixed},
   		xlabel = {Time (ms)},
   		ylabel = {$\theta_{13400}$ (rad)},
   		label style = {font=\scriptsize},   	
   		grid = major,
		ytick = {-0.16, -0.02, 0.13, 0.27},
		ymin = -0.22, ymax = 0.34,
   		tick label style = {font=\scriptsize},
   		set layers
   		]
   		
   		\begin{pgfonlayer}{axis background}
			\path[fill=gray!15] (axis cs:0,  -0.22) rectangle (axis cs:20, 0.34);
			\path[fill=gray!5]  (axis cs:20, -0.22) rectangle (axis cs:40, 0.34);
			\path[fill=gray!15] (axis cs:40, -0.22) rectangle (axis cs:60, 0.34);
			\path[fill=gray!5]  (axis cs:60, -0.22) rectangle (axis cs:80, 0.34); 
   		\end{pgfonlayer}
   		
   		\addplot [red, mark=*, only marks, mark indices={1, 2, 3, 4, 5, 6, 7, 8}, mark options={solid}, mark size=1.5pt] 
   		coordinates {
   			(0,  -0.137912) (20,  -0.137912) 		
   			(20, -0.154070) (40, -0.154070) 
   			(40, -0.170680) (60, -0.170680) 
   			(60,  0.285875) (80,  0.285875)
   		};
   	
   		\addplot [blue,  mark=*, mark size=0.6pt] 
   		coordinates {	
   			(0, -0.137948) (2.63, -0.137948) (4.85, -0.137948) (7.08, -0.137948) (9.30, -0.137948) (11.50, -0.137948)
   			(13.69, -0.137948) (15.89, -0.137948) (18.08, -0.137948) (20.28, -0.137948)
   			(22.47, -0.147769) (24.69, -0.153928) (26.79, -0.154123) (28.86, -0.154123)
			(31.19, -0.154123) (33.33, -0.154123) (35.36, -0.154123) (37.48, -0.154123) (39.53, -0.154123) (41.64, -0.154123)
   			(43.72, -0.168401) (46.17, -0.170758) (48.33, -0.170573) (50.50,-0.170573) 
   			(52.67, -0.170573) (54.83, -0.170573) (57.00, -0.170573) (59.17, -0.170573) (61.33, -0.170573)
   			(63.50, 0.285937) (65.67, 0.285827) (67.83, 0.285912)
   			(70.00, 0.285912) (72.16, 0.285912) (74.33, 0.285912) (76.50, 0.285912) (78.66, 0.285912) (80.83, 0.285912) 	
   		};
   	\end{axis}
   	
	\end{tikzpicture}}
	\end{tabular}
	\caption{
		Voltage magnitude and angle at bus 430 (subfigure a) and bus 13400 (subfigure b), estimated by the fusion-based GBP algorithm 
		under asynchronous PMU updates for the 13659-bus power system. The red circles indicate transitions between power-system operating
		conditions and the corresponding exact values, while the blue dots denote the successive GBP estimates, illustrating how the algorithm 
		detects and tracks the new operating state over time.
	}
	\label{plot4}
\end{figure} 

The estimated voltage magnitude and angle at buses 430 and 13400 are shown in \figurename~\ref{plot4}, where the blue markers correspond to the GBP estimates obtained at successive iterations, while the red markers denote the exact state values sought by the algorithm. The initial delay, during which the GBP estimate still reflects the previous operating condition even after the system state has changed, arises because the algorithm is already in the middle of an iteration and cannot incorporate newly arriving measurements until the next update cycle. After that point, when $60\%$ of the phasor measurements are refreshed, the accuracy with which the change is captured depends on the updated phasor measurements available at that iteration. In particular, \figurename~\ref{plot4b} shows that the available updated measurements are sufficient to recover almost exactly the new state at bus 13400 within a single iteration. By contrast, \figurename~\ref{plot4a} shows that the available updated measurements are not sufficient to recover the exact new state at bus 430 immediately. Nevertheless, the GBP estimate reacts at once, clearly indicating that the system has departed from the previous operating condition. In the following iterations, as additional phasor measurements are refreshed, the estimate rapidly aligns with the new operating state. These results demonstrate that the continuously running GBP algorithm can support near real-time detection of operating-condition changes while rapidly recovering an accurate estimate of the new system state under asynchronous measurement updates.

\section{Conclusions}
A vectorized GBP framework for PMU-based SE is developed over factor graphs through multivariate and fusion-based formulations. Compared to the scalar formulation, the multivariate formulation captures measurement correlations and reduces graph complexity. The fusion-based formulation further simplifies the graph by combining measurements associated with the same set of state variables, which reduces redundancy and improves convergence behavior relative to both the scalar and multivariate formulations. The fusion-based formulation also demonstrates that the proposed framework can accommodate alternative aggregation schemes, such as PMU-oriented fusion of current phasor measurements on branches incident to the same bus. Given this flexibility, the fusion-based GBP also achieves the best overall performance, providing accurate estimates after only a few iterations while remaining robust under correlated measurement errors, increasing redundancy, growing system size, and asynchronous PMU updates. At the same time, the fusion-based formulation leads to a factor graph structure that more closely reflects the underlying bus/branch model and the associated measurement topology, which further reinforces its practical relevance. These findings indicate strong potential for distributed and near real-time monitoring of modern power systems.

\section*{Appendix A: Canonical Form of the GBP Algorithm}
For efficient implementation, it is convenient to express all Gaussian messages in canonical (information) form. In this representation, instead of exchanging the mean and precision pairs $\mathbf{z}_{f_k \to \mathbf{x}_i}$, $\mathbf{\Lambda}_{f_k \to \mathbf{x}_i}$ and $\mathbf{z}_{\mathbf{x}_i \to f_k}$, $\mathbf{\Lambda}_{\mathbf{x}_i \to f_k}$, as in the moment form, the messages are expressed in terms of the information and precision pairs $\boldsymbol{\eta}_{f_k \to \mathbf{x}_i}$, $\mathbf{\Lambda}_{f_k \to \mathbf{x}_i}$ and $\boldsymbol{\eta}_{\mathbf{x}_i \to f_k}$, $\mathbf{\Lambda}_{\mathbf{x}_i \to f_k}$. The canonical form is obtained from the moment form by expressing the means as:
\begin{subequations}
	\begin{align}
		\mathbf{z}_{f_k \to \mathbf{x}_i} &= \mathbf{\Lambda}_{f_k \to \mathbf{x}_i}^{-1} \boldsymbol{\eta}_{f_k \to \mathbf{x}_i} \\
		\mathbf{z}_{\mathbf{x}_i \to f_k} &= \mathbf{\Lambda}_{\mathbf{x}_i \to f_k}^{-1} \boldsymbol{\eta}_{\mathbf{x}_i \to f_k}.
	\end{align}
\end{subequations}
Within this representation, the GBP algorithm can be reformulated in a way that avoids certain matrix-vector products and explicit inversions appearing in the moment form. 

In particular, the message $\mu_{f_k \to \mathbf{x}_i}(\mathbf{x}_i)$ from a pairwise factor node to a variable node is given by:
\begin{subequations}
    \begin{align}
    		\scalebox{0.95}{$\mathbf{\Lambda}_{f_k \to \mathbf{x}_i}$} &= 
        \scalebox{0.95}{$\mathbf{H}_{f_k,\mathbf{x}_i}^T \mathbf{V}_{f_k}^{-1} \mathbf{H}_{f_k,\mathbf{x}_i}$} \\
		\scalebox{0.95}{$\boldsymbol{\eta}_{f_k \to \mathbf{x}_i}$} &= 
        \scalebox{0.95}{$\mathbf{H}_{f_k,\mathbf{x}_i}^T \mathbf{V}_{f_k}^{-1} \left(\mathbf{z}_{f_k} - 
		\mathbf{H}_{f_k,\mathbf{x}_j} \mathbf{\Lambda}_{\mathbf{x}_j \to f_k}^{-1} \boldsymbol{\eta}_{\mathbf{x}_j \to f_k}\right).$}
    \end{align}
\end{subequations}
Compared to the moment form, the expression for the precision matrix coincides with \eqref{BP_fv_var}. In contrast to \eqref{fused_mean_message}, the inversion of $\mathbf{\Lambda}_{f_k \to \mathbf{x}_i}$ is avoided, while the inverse of $\mathbf{\Lambda}_{\mathbf{x}_j \to f_k}$ is introduced. Consequently, the overall computational complexity of this message update remains unchanged.

The message $\mu_{\mathbf{x}_i \to f_k }(\mathbf{x}_i)$ from a variable node to a pairwise factor node is given by:
\begin{subequations}
	\begin{align}
		\mathbf{\Lambda}_{\mathbf{x}_i \to f_k} &= 
		\sum_{f_a \in \mathcal{F}_i \setminus \{f_k\}} \mathbf{\Lambda}_{f_a \to \mathbf{x}_i} \\
		\boldsymbol{\eta}_{\mathbf{x}_i \to f_k} &= 
		\sum_{f_a \in \mathcal{F}_i \setminus \{f_k\}} \boldsymbol{\eta}_{f_a \to \mathbf{x}_i}.
	\end{align}
\end{subequations}
Compared to the moment form, the expression for the precision matrix coincides with \eqref{BP_vf_var}. However, unlike \eqref{BP_vf_mean}, the update avoids explicit inversion of $\mathbf{\Lambda}_{\mathbf{x}_i \to f_k}$. Moreover, the matrix-vector products $\mathbf{\Lambda}_{f_a \to \mathbf{x}_i} \mathbf{z}_{f_a \to \mathbf{x}_i}$ are eliminated, and the update reduces to simple summation of incoming precision matrices and information vectors.

The marginal distribution at variable node $\mathbf{x}_i$ is obtained as:
\begin{subequations}
	\begin{align}
		\mathbf{\Lambda}_{\mathbf{x}_i} &= 
		\sum_{f_a \in \mathcal{F}_i} \mathbf{\Lambda}_{f_a \to \mathbf{x}_i} \\
		\hat{\mathbf{x}}_i &= 
		\mathbf{\Lambda}_{\mathbf{x}_i}^{-1} \sum_{f_a \in \mathcal{F}_i} \boldsymbol{\eta}_{f_a \to \mathbf{x}_i}.
	\end{align}
\end{subequations}
Compared to the moment form in \eqref{BP_marginal_mean_var}, the precision matrix is computed in the same way, while the marginal estimate is obtained from the marginal precision matrix and the sum of incoming information vectors, avoiding the intermediate matrix-vector products $\mathbf{\Lambda}_{f_a \to \mathbf{x}_i} \mathbf{z}_{f_a \to \mathbf{x}_i}$.

\section*{Appendix B: Broadcast Form of the GBP Algorithm}
In addition to the canonical representation, the broadcast scheme provides a second approach for efficient implementation of the GBP algorithm. Instead of computing each message by summing all incoming messages except the one associated with the target edge, and repeating this operation for every target edge, the broadcast approach first aggregates all incoming messages and then obtains each message by subtracting the contribution corresponding to the target edge. This approach affects only the messages from variable nodes to factor nodes, while messages from factor nodes to variable nodes remain unchanged, since each factor node receives only a single incoming message from the corresponding variable node. As a result, the computational complexity of message updates at variable nodes is reduced from $\mathcal{O}(|\mathcal{F}_i|^2)$ to $\mathcal{O}(|\mathcal{F}_i|)$, where $|\mathcal{F}_i|$ denotes the number of neighboring factor nodes of variable node $\mathbf{x}_i$.

The message $\mu_{\mathbf{x}_i \to f_k }(\mathbf{x}_i)$ from a variable node to a pairwise factor node is given by:
\begin{subequations}
	\begin{align}
		\mathbf{\Lambda}_{\mathbf{x}_i \to f_k} &= \mathbf{\Lambda}_{\mathbf{x}_i} - \mathbf{\Lambda}_{f_k \to \mathbf{x}_i} \\
		\boldsymbol{\eta}_{\mathbf{x}_i \to f_k} &= \boldsymbol{\eta}_{\mathbf{x}_i} - \boldsymbol{\eta}_{f_k \to \mathbf{x}_i},
	\end{align}
\end{subequations}
where:
\begin{subequations}
	\begin{align}
		\mathbf{\Lambda}_{\mathbf{x}_i} &= \sum_{f_a \in \mathcal{F}_i} \mathbf{\Lambda}_{f_a \to \mathbf{x}_i} \\
		\boldsymbol{\eta}_{\mathbf{x}_i} &= \sum_{f_a \in \mathcal{F}_i} \boldsymbol{\eta}_{f_a \to \mathbf{x}_i}.
	\end{align}
\end{subequations}
The aggregate quantities $\mathbf{\Lambda}_{\mathbf{x}_i}$ and $\boldsymbol{\eta}_{\mathbf{x}_i}$ are identical for all messages sent from variable node $\mathbf{x}_i$ and are therefore computed only once. Moreover, the aggregated quantities $\mathbf{\Lambda}_{\mathbf{x}_i}$ and $\boldsymbol{\eta}_{\mathbf{x}_i}$ define the local canonical parameters at variable node $\mathbf{x}_i$ and can therefore be reused when forming the corresponding marginal belief.

\bibliographystyle{IEEEtran}
\bibliography{cite}

\end{document}